\documentclass[aps,pra,twocolumn,nofootinbib,showpacs]{revtex4}

\usepackage{graphicx}

\begin{document}

\title{Planckian relaxation delusion in metals}

\author{M.V. Sadovskii}

\affiliation{Institute for Electrophysics, Russian Academy of Sciences, 
Ural Branch, Ekaterinburg 620016, Russia
\\
E-mail: sadovski@iep.uran.ru}

\begin{abstract}

We present a critical review of recent attempts to introduce the new quantum
(``Planckian'') limit for the temperature dependence of inelastic scattering 
rate of electrons in metals. We briefly discuss the main experimental facts
and some simple theoretical models explaining the linear in temperature growth 
of resistivity (starting from very low temperatures) in superconducting 
cuprates and some similar systems. There is no commonly accepted theoretical
explanation of such behavior up to now. We also discuss the known quantum
limits for electrical conductivity (resistance). It is shown that the universal
Planckian limit for the inelastic relaxation rate proposed in some papers 
is a kind of delusion related to a certain procedure to represent the
experimental data.

\end{abstract}


\maketitle

\tableofcontents

\newpage

\section{Introduction}

Linear in temperature ($T$ -- linear) growth of electrical resistivity of
cuprates and some other correlated systems in the wide temperature region
from pretty low to rather high temperatures remains among the most challenging
problems of the physics of high -- temperature superconductors for many years. 
Observed in cuprates since the early experiments \cite{Batl,Iye} it became
one of the marked properties of these compounds in normal phase, starting in
optimally doped systems from the temperature of superconducting transition
$T_c$ and continuing up to highest temperatures achievable before destruction 
the samples. Later the similar behavior of resistivity was observed in some
other similar systems. By itself, the $T$ -- linear growth of resistivity in
metals is not at all surprising, it is observed practically always, though at 
high enough temperatures $T>\Theta_D/5$, where $\Theta_D$ -- is Debye
temperature, which is usually some hundreds of degrees. What is surprising
is the fact, that the $T$ -- linear growth in cuprates takes place, starting 
from significantly lower temperatures. The commonly accepted explanation of 
such behavior is still lacking, though during these years a number of theoretical
models were proposed, claiming such explanation. But this is not the aim of the
current work.

The thing is that recently a number of interesting papers appeared 
\cite{Bruin,Legros}, where after rather detailed analysis of a wide experimental
material on many compounds, it was shown that in the $T$ -- linear region of
resistivity, the scattering rate of electrons (inverse relaxation time) is
pretty accurately described by the dependence
$\Gamma(T)=1/\tau=\alpha k_BT/\hbar$, where $\alpha\sim 1$ and is weakly
varying from one material to another. In particular, for systems in the
vicinity of quantum critical points (on the phase diagram of cuprates and
some similar systems) the value of $\alpha$ is in the interval of
0.7 -- 1.1, and is seemingly universal and independent of the strength of
interaction, leading to relaxation of electronic current. More so, it was
discovered that the similar dependence gives rather good description of the
data for a number of usual metals  (like Cu, Au, Al, Ag, Pb, Nb etc.) in
the $T$ -- linear region of their resistivity. In this case the values of
$\alpha$ belong to noticeably wider interval from 0.7 to 2.7 \cite{Bruin,Legros}. 
In connection with these (and similar) results the concept of the universal
(independent of interaction strength) ``Planckian'' {\em upper} limit of
scattering rate $\frac{1}{\tau_P}=\Gamma_P\sim\frac{k_BT}{\hbar}$ was 
introduced in Ref. \cite{Zaanen}.

Below we shall present a short review of the relevant experimental data and
discuss some theoretical models, imposing quantum limitations on resistivity of
metals, with the aim to understand the degree of validity of the concept of
``Planckian'' relaxation in metals.


\section{Experiments}

Temperature dependence of resistivity at low enough temperatures 
($T< 0.2\Theta_D\ll E_F$, $\Theta_D$ -- Debye temperature, $E_F$ -- Fermi
energy) is described by the following expression:
\begin{equation}
\rho(T)=\rho_0+\rho_T(T)=\rho_0+a_{ee}\left(\frac{k_BT}{E_F}\right)^2
+a_{ep}\left(\frac{T}{\Theta_D}\right)^5
\label{rhoT}
\end{equation}
where $\rho_0$ -- is residual resistivity at zero temperature due to impurity
scattering, the second term is the contribution electron -- electron
scattering, while the third one is the low temperature contribution from
electron -- phonon scattering, described by Bloch -- Gr\"uneisen theory \cite{Zim}.
For high enough temperatures $T> 0.2\Theta_D$ this theory gives:
\begin{equation}
\rho(T)-\rho_0=AT
\label{AT}
\end{equation}
This behavior of resistivity is clearly seen in the experiments, as shown in
Fig. \ref{B_G}, where we show the data for a number of simple metals \cite{Meiss}.
These results show that the temperature dependence of resistivity (conductivity)
is almost totally related to the processes of inelastic scattering by phonons.
In case of significant contribution of scattering by other collective 
excitations, e.g. spin fluctuations, we can write down in fact very similar
expressions.

From these data it is seen, that for temperatures $T>\Theta_D/5$ resistivity of
a metal grows linearly with temperature. In most metals $\Theta_D\sim$ 200--600K.
\begin{figure}
\includegraphics[clip=true,width=0.45\textwidth]{Bl_Gr.eps}
\caption{Temperature dependence of resistivity for typical metals.
\cite{Meiss}}
\label{B_G}
\end{figure}
At the same time, the critical temperature of superconducting transition in 
cuprates is commonly some dozens of degrees, so that $T_c\sim\Theta_D/5$.
Thus, for rather long time there was a hope \cite{Maks} that $T$ -- linear 
growth of resistivity in cuprates can be explained by the usual electron --
phonon scattering, taking into account that it takes place outside the region
of low temperature power -- like growth, which is masked by superconducting
transition. There were some cases of few exceptional samples with
low values of $T_c$, where $T$ -- linear growth of resistivity was observed
starting from anomalously low temperatures $T<10$K \cite{Fior}, but these
were rather seldom. There were no experiments in the normal phase at low
enough temperatures simply because the destruction of superconducting state in
typical cuprates required extremely high magnetic fields.

This situation changed in recent years after such experiments were 
successfully performed \cite{Bruin,Legros}. These works presented the data on
resistivity in a number of high -- temperature cuprate superconductors as well
as on some analogous systems in very high magnetic fields, suppressing
superconductivity. The detailed analysis of these experiments and data obtained
by other authors has shown that $T$ -- linear behavior of resistivity is
conserved in normal phase, in many cases, up to the lowest temperatures.

\begin{figure}
\includegraphics[clip=true,width=0.45\textwidth]{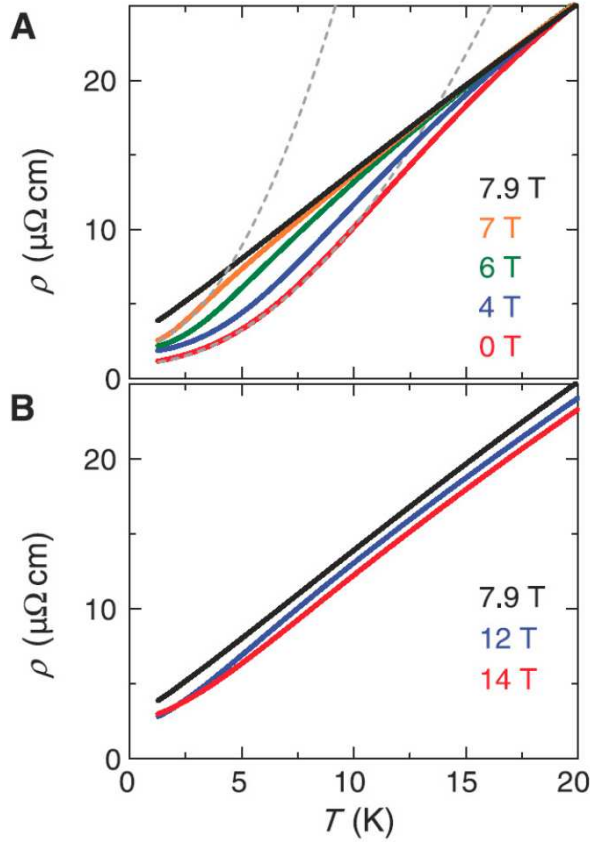}
\caption{(A) Resistivity $\rho$ as function of temperature $T$ of very pure
single crystal of Sr$_3$Ru$_2$O$_7$ in magnetic fields 0T (red line),
4T (blue), 6T (green), 7T (orange) and in critical field $H_c$=7.9T (black),
corresponding to quantum critical point ($H_c$,$T=0$).
Dashed lines -- a fit to $\rho_0+AT^2$ dependence at low temperatures,
which demonstrates the narrowing of the region of quadratic dependence on
temperature and crossover to linear dependence. \cite{Bruin}
(B) Temperature dependence of resistivity $\rho$ in fields $H_c$ (black),
12T (blue) and 14T (red). \cite{Bruin}}
\label{Linr_1}
\end{figure}

\begin{figure}
\includegraphics[clip=true,width=0.45\textwidth]{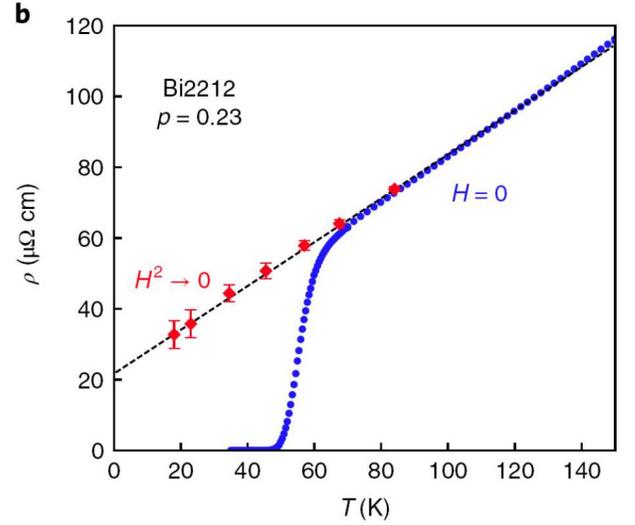}
\caption{Resistivity of Bi2212 thin film as function of temperature for
$H=$0 (blue curve). Red dots (and dashed line) show resistivity in normal phase
for $T<T_c$, obtained by suppressing superconductivity by external magnetic
field and extrapolation from $\rho=a+bH^2$ dependence. \cite{Legros}}
\label{Linr_2}
\end{figure}

\begin{figure}
\includegraphics[clip=true,width=0.45\textwidth]{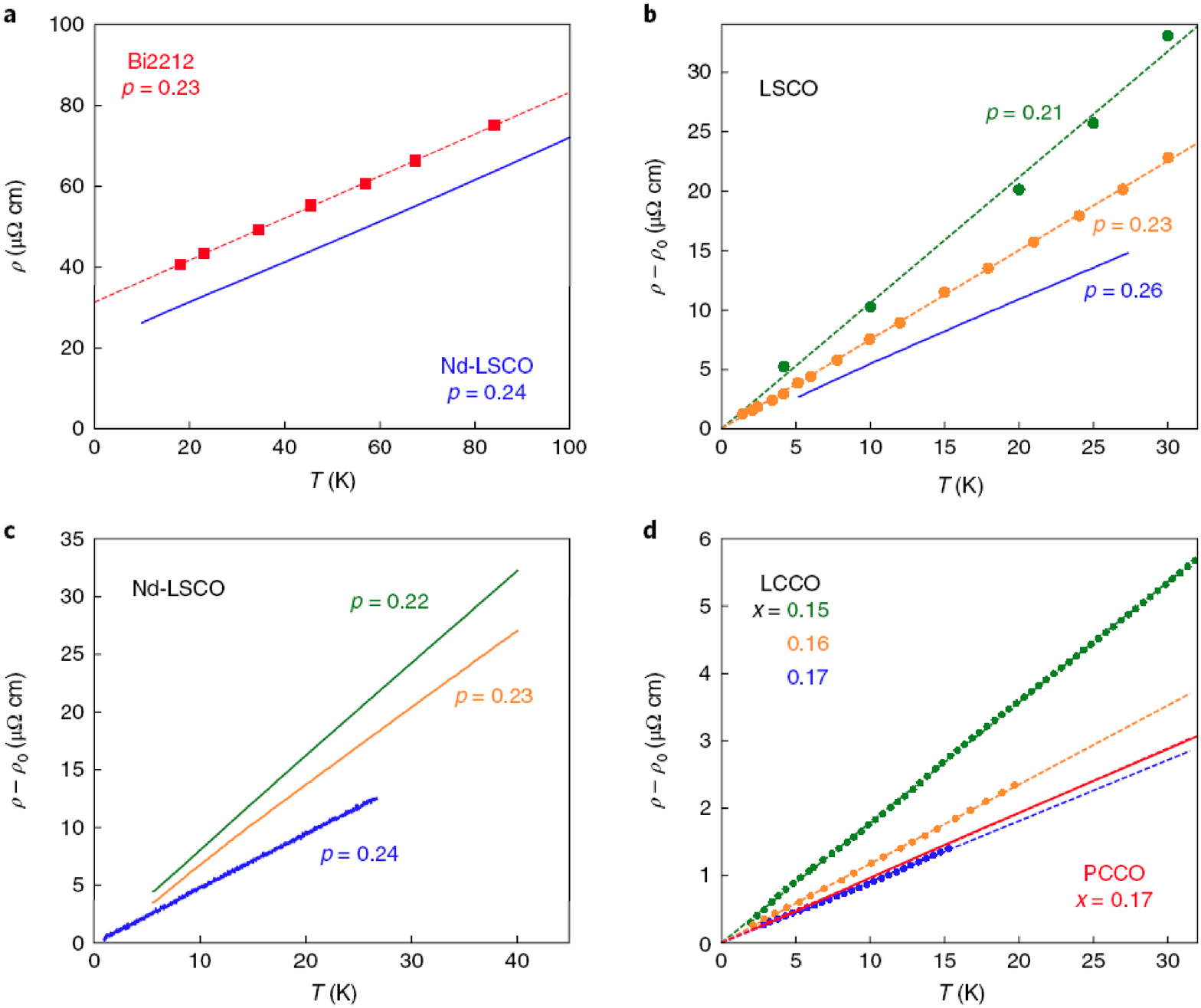}
\caption{Linear in temperature dependence of resistivity in overdoped cuprates,
showing the behavior of $\rho$ in highly -- conducting plane (i.e. $\rho_{ab}$) in
(a) Nd-LSCO for $p=0.24$ (blue line, $H=$16T) and Bi2212 for $p=$0.23 
(red symbols), $H=$55T); (b) $\rho(T)-\rho_0$ in LSCO for $p=$0.21 
(green symbols, $H=$48T)
and $p=$0.23 (orange, $H=$48T), $p=$0.26 (blue, $H=$18T); 
(c) $\rho(T)-\rho_0$
in  Nd-LSCO at $H=$33T for $p=$0.22 (green), 
$p=$0.23 (orange) and $p=$0.24 (blue); (d) $\rho(T)-\rho_0$ in  LCCO for 
$x=$0.16 (orange, $H=$6.5T) and
$x=$0.17 (blue, $H=$4T) and PCCO for $x=$0.17 (red, $H=$16T). \cite{Legros}}
\label{Linr_3}
\end{figure}

Typical examples of experimental data from Refs. \cite{Bruin,Legros} are shown
in Fig. \ref{Linr_1}, \ref{Linr_2}, \ref{Linr_3}. In these works, characterized
by the very detailed analysis of experimental situation, many additional data
can be found. These results aggravated further the question of the nature of
$T$ -- linear resistivity in systems under consideration. It should be stressed
from the very beginning, that this problems remains unsolved up to now and this
is not the aim of the present work.

For us the main interest now is the analysis of experimental data presented in
Refs. \cite{Bruin,Legros}, which allowed the authors to determine the 
temperature dependence of relaxation time $\tau$, from the values of resistivity
and to come to rather unexpected results and conclusions.

The main idea of the analysis performed in Refs. \cite{Bruin,Legros} was as
follows. Let us write down the Drude expression for conductivity as:
\begin{equation}
\sigma=\frac{ne^2\tau}{m^{\star}}=\frac{ne^2v_F\tau}{\hbar k_F}
\label{Drud}
\end{equation}
where $v_F$ is electron velocity at Fermi surface, $m^{\star}$ is an effective
mass, $p_F=m^{\star}v_{F}=\hbar k_F$ is Fermi momentum. Correspondingly for
resistivity we have\footnote{For shortness we write down here all expressions
for one-band model. In multiple band case we have to take into account 
contributions from all pockets of the Fermi surface. Appropriate expressions
can be found in Refs. \cite{Bruin,Legros}.}:
\begin{equation}
\rho=\frac{m^{\star}}{ne^2\tau}=\frac{\hbar k_F}{ne^2v_F\tau}
\label{soprot}
\end{equation}
The effective mass $m^{\star}$ or the value of $\hbar k_F/v_F$ 
in Refs. \cite{Bruin,Legros} were determined from electronic
contribution to specific heat, which can be measured at low temperatures, 
or from the measurements of de Haas -- van Alfen effect, which are also made
at low enough temperatures. These measurements in fact determine the value of
the effective mass $m^{\star}$ and Fermi wave -- vector $k_F$  as some average
values for each of the pockets of the Fermi surface.

Electron density $n$, entering these expressions can be calculated (for systems
of different dimensionality) as:
\begin{eqnarray}
n=\frac{2k_F}{\pi d_bd_c}\qquad \mbox{(quasi-one-dimensional case)}\\
n=\frac{k_F^2}{2\pi d}\qquad \mbox{(quasi-two-dimensional case)}\\
n=\frac{k_F^3}{3\pi^2}\qquad \mbox{(three-dimensional case)}
\label{density}
\end{eqnarray}
where $d_b$ and $d_c$ are the distances between adjacent conducting chains
(oriented along $a$-axis) in the directions of $b$ and $c$-axis in quasi --
one -- dimensional system, while $d$ is the distance between adjacent
conducting planes in quasi -- two -- dimensional case. 

If we express the temperature -- dependent part of resistivity as in (\ref{AT})
and introduce the $T$ -- linear (Planckian) relaxation rate as:
\begin{equation}
\Gamma=\frac{1}{\tau}=\alpha \frac{k_BT}{\hbar};\quad \alpha\sim const
\label{Plk_scatt}
\end{equation}
we immediately obtain:
\begin{eqnarray}
A=\frac{d\rho}{dT}=\alpha\frac{k_B}{\hbar}\frac{m^{\star}}{ne^2}\quad\mbox{or}\\
\frac{ne^2}{k_Bk_F}\frac{d\rho}{dT}=\alpha\frac{m^{\star}}{\hbar k_F}=
\frac{\alpha}{v_F}
\label{Acoef}
\end{eqnarray}
which gives working formulae to represent experimental data in $T$ -- linear
region of resistivity \cite{Bruin,Legros}.

In Refs. \cite{Bruin,Legros} the very detailed analysis of experimental data
was performed for rather wide set of systems (compounds) with very different
electronic structures, from high -- temperature superconducting copper oxides
and iron based superconductors to organic metals like (TMTSF)$_2$PF$_6$,
through to compounds like Sr$_3$Ru$_2$O$_7$, CeCoIn$_5$, UPt$_3$, CeRu$_2$Si$_2$, 
where the $T$ -- linear growth of resistivity is observed.

It was discovered, that for all these quite different systems, the experimental
data for thus defined relaxation rate are well described by dependence like
(\ref{Plk_scatt}) and the value of $\alpha$ for majority of these systems
belong to the interval 0.7 -- 1.1, and seems to be universal (independent of
peculiarities of electronic spectrum or the strength of interaction, leading to
scattering of electrons). More so, it was also shown, that the similar
dependence is appropriately describing also the data for a number of usual
metals (Cu, Au, Al, Ag, Pb, Nb), though the values of $\alpha$ for them belong
to a wider interval from 0.7 to 2.8 \cite{Bruin,Legros}.

\begin{figure}
\includegraphics[clip=true,width=0.45\textwidth]{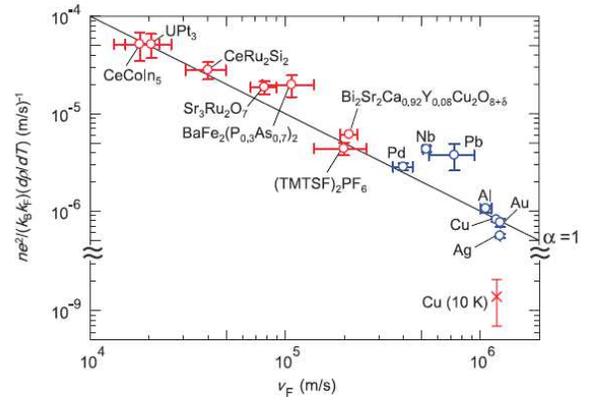}
\caption{The general picture for different metals (different Fermi velocities
$v_F$), demonstrating the the $T$ -- linear growth of resistivity.
These metals include heavy -- fermion systems, oxides, pnictides and organic
metals, where the $T$ -- linear growth of resistivity at low temperatures
appears after the application of magnetic field, change of the chemical
composition or external hydrostatic pressure. Analogous data represent the
situation in common metals with $T$ -- linear behavior of resistivity at
high enough temperatures (blue symbols), while for lower temperatures their
resistivity is orders of magnitude smaller (appropriate value for Cu at
$T=$10K is shown in right lower corner). Straight line, denoted by
$\alpha=1$, corresponds to $(\tau T)^{-1}=k_B/\hbar$. \cite{Legros}}
\label{LinPl}
\end{figure}

In the Table given below we show, as an example, the experimental values of
some parameters under discussion, determined in Ref. \cite{Legros}, for a
number of quasi -- two -- dimensional systems (hole doped and electron doped
cuprates, organics).

\begin{figure*}

\footnotesize

\begin{tabular}{cccccc}
\hline

Compound & Doping & $n$(10$^{27}$m$^{-3}$) & $m^{\star}/m_0$ & $A/d$($\Omega$ K$^{-1}$) & $\alpha$ \\
\hline
Bi$_2$Sr$_2$CaCu$_2$O$_{8-\delta}$ (Bi2212)    &  $p=$0.23  &   6.8  &  8.4$\pm$1.6  &  8.0$\pm$0.9 & 1.1$\pm$0.3 \\
Bi$_2$Sr$_2$CuO$_{6+\delta}$ (Bi2201)   &  $p=$0.4   &   3.5  &  7$\pm$1.5    &  8$\pm$2     & 1.0$\pm$0.4\\
La$_{2-x}$Sr$_x$CuO$_4$ (LSCO)     &  $p=$0.26  &   7.8  &  9.8$\pm$1.7  &  8.2$\pm$1.0 & 0.9$\pm$0.3\\
La$_{1.6-x}$Nd$_{0.4}$Sr$_x$CuO$_4$ (Nd-LSCO)  &  $p=$0.24  &   7.9  &  12$\pm$4     &  7.4$\pm$0.8 & 0.7$\pm$0.4\\
Pr$_{2-x}$Ce$_x$CuO$_{4\pm\delta}$ (PCCO)     &  $x=$0.17  &   8.8  &  2.4$\pm$0.1  &  1.7$\pm$0.3 & 0.8$\pm$0.2\\
La$_{2-x}Ce$$_x$Cu)$_4$ (LCCO)     &  $x=0.15$  &   9.0  &  3.0$\pm$0.3  &  3.0$\pm$0.45& 1.2$\pm$0.3\\
(TMTSF)$_2$PF$_6$ (TMTSF)    &  $P=$11kbar&   1.4  &  1.15$\pm$0.2 &  2.8$\pm$0.45& 1.0$\pm$0.3\\
\hline

\end{tabular}

\normalsize

\end{figure*}

In Refs. \cite{Bruin,Legros} further details can be found on similar data 
for all systems mentioned above.

All this is nicely illustrated by consolidated graph shown in Fig. \ref{LinPl}, 
where data are shown in logarithmic scale representing Eq. (\ref{Acoef}). 
These results seem to be quite nontrivial and apparently confirm the concept of
universal ``Planckian'' mechanism of electronic relaxation in metals, which was
introduced earlier \cite{Zaanen} and applied to cuprates physics.
More so, the value of $\alpha\sim 1$, which is observed  for such a wide
set of materials with very different electronic spectra and quite different
Fermi surfaces, suggests the idea that (\ref{Plk_scatt}) is actually the
universal quantum {\em upper} limit for inelastic (temperature -- dependent) 
relaxation rate for electrons in metals.

To explain such temperature behavior of resistivity for so different systems,
from the lowest temperatures, a number of complicated theoretical models was
proposed recently \cite{Khod,Khodel,Sachdev,Vol}, including some very exotic,
taken from the physics of black holes, cosmology and superstring theory
(cf. \cite{Zan,Hart,Herz,Hartn}). Below we shall limit ourselves to a simple
analysis based on the traditional approaches of quantum theory of solids.


\section{Quantum estimates for resistivity of metals.}

Let us remind some elementary theoretical estimates with respect to conductivity
(resistivity) of metals. Drude expressions for elastic and inelastic
scattering are written as:
\begin{eqnarray}
\sigma_0=\frac{ne^2}{m}\tau_{0}; \qquad \rho_0=\frac{m}{ne^2}\frac{1}{\tau_0}\\
\sigma_{i}(T)=\frac{ne^2}{m}\tau(T); \qquad \rho_{i}(T)=\frac{m}{ne^2}
\frac{1}{\tau(T)}
\label{sigma}
\end{eqnarray}
where $\tau_{0}$ -- is the mean free time due to elastic impurity scattering, 
determining the residual resistivity, and $\tau(T)$ is relaxation time due to
inelastic scattering by phonons (or some other collective excitations), or
due to electron -- electron scatterings.
Mass $m$ here is always understood as free electron mass (band structure mass
in a solid!), which {\em does not include} corrections due to electron --
electron or electron -- phonon interactions (see more details below).
Then, assuming the additive contributions of different scattering mechanisms
(Matissen rule), the total resistivity is written as:
\begin{equation}
\rho(T)=\frac{m}{ne^2}\left(\frac{1}{\tau_{0}}+
\frac{1}{\tau(T)}\right)\equiv\frac{m}{ne^2}(\Gamma_{0}+\Gamma(T))
\label{sigmainv}
\end{equation}
where we have introduced the appropriate scattering rates:
\begin{eqnarray}
\Gamma_{0}=\frac{1}{\tau_0}\\
\Gamma(T)=\frac{1}{\tau(T)}
\label{gammas}
\end{eqnarray}
In general theory of interacting fermions (electrons) by the order of magnitude
we have $\Gamma\sim Im \Sigma$, where $\Sigma$ is an electron self -- energy,
taking into account all the relevant interactions. Consistent approach to
calculation of conductivity (resistivity) requires, of course, the treatment of
a full two -- particle Green's function \cite{Diagram}.

\subsubsection{Ioffe -- Regel limit}

The most prominent quantum limitation for conductivity (resistivity) of metals
is the Ioffe -- Regel limit \cite{IR}, which is relevant to strongly disordered
systems. For $T=0$ we have:
\begin{equation}
\sigma_0=\frac{ne^2}{m}\tau_{0}=\frac{ne^2}{p_F}l
\label{sig_0}
\end{equation}
where $l=v_F\tau_0$ is the mean free path.
The usual kinetic theory is valid for $p_Fl/\hbar\gg 1$ or
$E_F\tau_{0}/\hbar\gg 1$.
Taking into account $n=\frac{p_F^3}{3\pi^2\hbar^3}$ we get the estimate for
conductivity in the Ioffe -- Regel limit as:
$p_Fl_{el}/\hbar\sim 1$ or $E_F\tau_{0}/\hbar\sim 1$:
\begin{equation}
\sigma_{IR}=\frac{e^2p_F}{3\pi^2\hbar^2}\sim\frac{e^2}{3\pi^2\hbar a}
\label{sigma_min}
\end{equation}
where $a\sim n^{-3}$ is the average distance between electrons. 
For typical metallic densities $a$ is of order of interatomic distance 
(lattice constant). In this case (for $n\sim$10$^{23}$ cm$^{-3}$) corresponding
resistivity $\rho_{IR}=\sigma_{IR}^{-1}\sim$150--300 $\mu\Omega$ cm. 
For majority of usual (``good'') metals, e.g. for Cu, $\rho(T=300$K)$\sim$
1 $\mu\Omega$ cm, so that this limit remains unachievable even at highest
temperatures, not exceeding the melting temperature. However, this is not so
in strongly disordered (highly -- resistive) alloys, where the resistivity can
approach the Ioffe -- Regel limit.

\footnotesize

An estimate quite close to (\ref{sigma_min}) was also proposed by Mott for the
so called ``minimal metallic conductivity'' $\sigma_c$, at which (achieved as
disorder grows e.g. mean free path being reduced) a discontinuous metal --
insulator transition takes place at  $T=0$ (Anderson localization) \cite{Mott}. 
In fact, as it is shown in scaling theory of localization \cite{AALR,Sad80}, 
this transition is continuous and $\sigma_c$ play a role of characteristic
conductivity scale in the vicinity of Anderson transition
($2<d<4$ -- space dimensionality):
\begin{equation}
\sigma\sim\sigma_c\left(\frac{l-l_c}{l_c}\right)^{(d-2)\nu}
\label{sigma_A}
\end{equation}
where $\sigma_c\sim {e^2}/{\hbar a^{d-2}}$ and the critical value of the mean
free path can be estimated from $p_Fl_c/\hbar \sim 1$, i.e. $l_c\sim a$.
Here we introduced the critical exponent of localization length, which in
self -- consistent theory of localization \cite{Diagram,VW} is given by:
\begin{equation}
\nu=\frac{1}{d-2}
\label{indRloc}
\end{equation}
so that $\nu=1$ for $d=3$. The modern numerical calculations of Anderson
transition give the values of $\nu$ in the interval 1.5--1.8 \cite{Markos}. 
For us not these details are important, but the fact of continuous nature
of this transition, and  nonexistense of any critical level of conductivity 
(resistivity).

Strictly speaking these estimates are valid only for $T=0$, when Anderson
transition is a well -- defined quantum phase transition in a system of 
noninteracting electrons. At finite temperatures and with the account of
electron -- electron interactions the situation is much more complicated and 
we shall not discuss it here.

\normalsize

In most cases the growth of resistivity with temperature slows down and
saturates as resistivity approaches the Ioffe -- Regel limit \cite{Gant}.
For such strongly disordered metals In such strongly disordered metals
(highly -- resistive alloys) an {\em empirical} Mooij rule is at work -- after
the achievement of resistivity level (at low temperatures) of the order of
$\rho_{IR}$, the temperature dependence of resistivity becomes very weak and
in a wide temperature interval from low to room temperatures and even higher
it is often observed that the temperature coefficient of resistivity becomes
{\em negative} \cite{Mooij,Gant}. This fact up to now does not have a commonly
accepted explanation.

There is a common belief, that this is not so in  ``strange'' metals like
high -- temperature superconducting copper oxides  \cite{Gur,Tak}
(in the region of optimal doping) and in a number of other systems \cite{Huss},
there the $T$ -- linear growth of resistivity continues even after overpassing
the values of the order of $\rho_{IR}=\sigma_{IR}^{-1}\sim$ 100--300 $\mu\Omega$ cm, 
up to highest possible temperatures $T\sim$1000 K. However, here we always face
the problem of the correct estimate of resistivity in Ioffe -- Regel limit.
From the estimates given above it becomes clear that it depends significantly
on concentration of current carriers (which in cuprates is much lower, than
in usual metals), so that e.g. for $n\sim$10$^{21}$ cm${^-3}$ we obtain 
$\rho_{IR}\sim$1--10 m$\Omega$ cm. In recent experiments of weakly doped cuprates
\cite{Greene} it was clearly shown, that their resistivity saturates at
$\rho_{IR}\sim$3--5 mOhm cm, in complete accordance with concentration of
carriers, obtained from the measurements of Hall effect (at $T\sim$300 K).
Thus it is quite possible, that in the experiments cited above on optimally
doped cuprates the correct value of Ioffe -- Regel limit was simply not
achieved up to highest possible temperatures (before the destruction of samples).

However, our main interest in the following will be related to resistivity of
``pure'' enough metals in $T$ -- linear region.


\subsubsection{Planckian relaxation}

The idea of Planckian relaxation mechanism in metal at high enough temperatures 
seems to be very attractive. Let us give some elementary arguments making it
seemingly justified and based upon quite general quantum mechanical estimates
using uncertainty principle \cite{Hart}. At finite temperatures $T>0$ different
processes of inelastic scattering (electron -- phonons, interaction with
spin -- fluctuations etc.) are at work. Precisely these processes are responsible
for establishing thermodynamic equilibrium in electronic system -- the Fermi
distribution. In a system of interacting particles at finite temperatures 
the particle (electron) distribution function is qualitatively of the same
form  \cite{Diagram}. Conductivity of a metal (degenerate case) is determined
by electron distribution in a layer $\sim k_BT$ around the Fermi level 
(chemical potential).
 Let us make an elementary estimate using time -- energy 
uncertainty relation:
\begin{equation}
\Delta E\tau > \hbar
\label{Heis}
\end{equation}
where $\tau$ is the life -- time of a quantum state, while $\Delta E$ is the
uncertainty of its energy. In our case $\tau=\tau(T)$ and it seems natural
to take $\Delta E\sim k_BT$, so that we immediately obtain an estimate:
\begin{equation}
\Gamma(T)=\frac{1}{\tau(T)}<\alpha\frac{k_BT}{\hbar}
\equiv\Gamma_{P}=\frac{1}{\tau_{P}}
\label{Planck}
\end{equation}
where $\alpha\sim 1$. Now it is evident that according to this elementary 
estimate the Planckian relaxation defines an {\em upper} limit for resistivity
due to inelastic scatterings: 
\begin{equation}
\rho(T)-\rho_0=\frac{m}{ne^2}\Gamma(T)<\frac{m}{ne^2}\alpha\frac{k_BT}{\hbar}
\equiv \rho_{P}(T)
\label{R_Planck}
\end{equation}
Obviously, this estimate is of rather speculative nature for the system of
many, in general, strongly  interacting particles, but it correlates well with
the results of experiments described above.

\begin{figure}
\includegraphics[clip=true,width=0.45\textwidth]{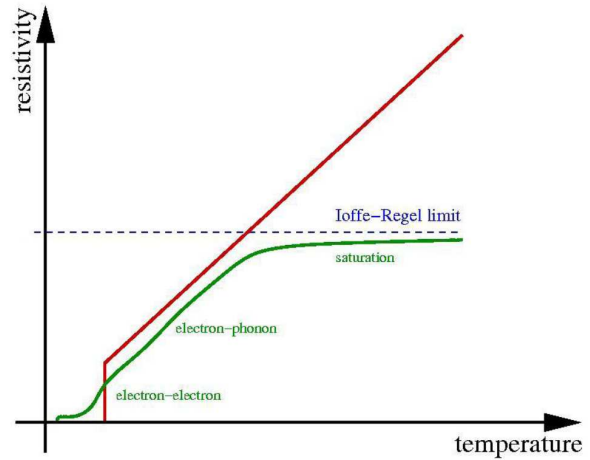}
\caption{``Planckian'' limit of electrical resistivity.
\cite{Zaane}
Red line -- the Planckian limit, green line -- the usual temperature 
dependence of resistivity in metals.}
\label{Zaan}
\end{figure}

If there exists such universal upper limit for relaxation rate, the qualitative 
picture of the temperature dependence of resistivity of metals can be suggested
as shown in Fig. \ref{Zaan} \cite{Zaane}. The idea here is that in ``usual''
metals all the temperature dependence of resistivity develops below the Ioffe --
Regel limit, while the ``Planckian'' limit is achieved in unusual systems like
HTSC -- cuprates, which are near the quantum critical point and where the Ioffe --
Regel limit may be surpassed with the growth of the temperature (however see
Ref. \cite{Greene}).

From the estimates given above it is easy to derive the ratio of ``Planckian''
resistivity to that of Ioffe -- Regel:\cite{Hart}
\begin{equation}
\frac{\rho_{P}(T)}{\rho_{IR}}\sim \frac{k_BT}{E_F}
\label{Plk_IR}
\end{equation}
so that this limit may be exceeded for $k_BT>E_F$, which can be realized
experimentally only in systems with low enough values of $E_F$
(low carrier concentration), such as copper oxides or in multiple band
systems (with several pockets of the Fermi surface). In any case, according to 
this picture resistivity does not exceed the upper limit, determined by the
``Planckian'' relaxation rate, which is achieved in ``strange'' metals. 
It is quite surprising that the experimental data quoted above seem to confirm
the achievement of this limit for many (!) metals, including some quite ``usual''.

\subsubsection{Electron -- phonon interaction}

Consider the most important for the theory of metals case -- the electron --
phonon interaction, which will be described within Eliashberg -- McMillan
theory, as the modern generalization of Bloch -- Gr\"uneisen theory \cite{Allen}.
Within this theory, the high -- temperature ($T>\Theta_D/5$) phonon contribution
to resistivity is given by \cite{Alle,Savr}:
\begin{equation}
\rho_{ph}(T)=\frac{4\pi mT}{ne^2\hbar}\int_{0}^{\infty}\frac{d\omega}{\omega}
\alpha_{tr}^2(\omega)F(\omega)=
\frac{2\pi mT}{ne^2\hbar}\lambda_{tr}
\label{RphonEMc}
\end{equation}
where  $\alpha_{tr}^2(\omega)F(\omega)$ is the transport Eliashberg -- McMillan
function \cite{Alle,Savr} ($F(\omega)$ is the phonon density of states),
determining the transport electron -- phonon coupling constant as:
\begin{equation}
\lambda_{tr}=2\int_{0}^{\infty}\frac{d\omega}{\omega}\alpha_{tr}^2(\omega)F(\omega)
\label{lamb_tr}
\end{equation}
For majority of metals we have $\alpha_{tr}^2\approx\alpha^2$, where \cite{Savr}
\begin{equation}
\lambda=2\int_{0}^{\infty}\frac{d\omega}{\omega}\alpha^2(\omega)F(\omega)
\label{lamb_McM}
\end{equation}
is Eliashberg -- McMillan coupling constant \cite{Allen}, determining the
temperature of superconducting transition. Then we get a simple estimate:
\begin{equation}
\rho_{ph}(T)\approx \frac{2\pi m T}{ne^2\hbar}\lambda\equiv 2\pi\lambda
\rho_{P}(T)
\label{RphMc}
\end{equation}
where in the last equality in the definition of $\rho_{P}(T)$ from 
(\ref{R_Planck}) we put $\alpha=1$. In fact the values of $\lambda > 1$ 
are not very rare in metals \cite{Allen} and it becomes clear, that even the
usual electron -- phonon interaction can easily {\em break} the Planckian
limit, so that $\rho_{ph}(T)>\rho_{P}(T)$ doses not seem exotic. This simple
example immediately casts certain doubts in universality of ``Planckian''
relaxation, though it seems that our arguments, based on uncertainty principle,
must work for any systems and interactions. However, this simple example just
contradicts it.


\section{Elementary model of scattering by quantum fluctuations in metals}

\subsubsection{Scattering by quantum fluctuations}

Here, following mainly Ref. \cite{Sad20}, we shall consider an elementary,
though realistic enough, model of electron scattering by quantum fluctuations,
which will allow us analyze our problem in rather general form.
Consider the usual Hamiltonian of electrons in a metal interacting with some
Bose -- like quantum fluctuations of some arbitrary nature ($N$ is her the total 
number of atoms in a crystal){\footnote
{Further on we shall use the units $\hbar=k_B=1$}:
\begin{equation}
H_{int}=\frac{1}{\sqrt N}\sum_{\bf pq}g_{\bf q}a^+_{\bf p+q}a_{\bf p}\rho_{\bf q}
\label{Hami}
\end{equation}
where we have used the standard notations of creation and annihilation 
operators of electrons, $\rho_{\bf q}$ is an operator of quantum fluctuation
of ``any kind'' (e.g. density of ions in a solid or collective excitations of
electronic subsystem, including spin excitations, though spin indices are
dropped for brevity). Let us introduce the appropriate (Matsubara time) Green's
function as \cite{Pin}:
\begin{equation}
F({\bf q},\tau)=-<T_{\tau}\rho_{\bf q}(\tau)\rho^+_{\bf q}(0)>
\label{GrFluc}
\end{equation}
Then we can write down the usual spectral representation for it as \cite{AGD}:
\begin{equation}
F({\bf q},i\omega_m)=\int_{-\infty}^{\infty}d\omega\frac{A({\bf q},\omega)}
{i\omega_m-\omega} 
\label{spectrrep}
\end{equation}
where $\omega_m=2\pi mT$, and the spectral density is defined as:
\begin{equation}
A({\bf q},\omega)=Z^{-1}\sum_{mn}e^{-\frac{E_n}{T}}|(\rho_{\bf q})_{nm}|^2
\left(1-e^{-\frac{\omega_{mn}}{T}}\right)\delta(\omega-\omega_{mn})
\label{specdens}
\end{equation}
where $\omega_{mn}=E_m-E_n$, $(\rho_{\bf q})_{nm}=<n|\rho_{\bf q}|m>
=(\rho^+_{\bf q})_{mn}$, and $m$, $n$ enumerate the exact many -- particle 
states of the system.

Dynamic structure factor of fluctuations is defined as \cite{Pin,PinNoz}:
\begin{equation}
S({\bf q},\omega)=Z^{-1}\sum_{mn}e^{-\frac{E_n}{T}}|(\rho_{\bf q})_{nm}|^2
\delta(\omega-\omega_{mn})
\label{dinstr}
\end{equation}
Comparing (\ref{specdens}) and (\ref{dinstr}) we obtain:
\begin{equation}
A({\bf q},\omega)=S({\bf q},\omega)\left[1-e^{-\frac{\omega}{T}}\right]
\label{AS}
\end{equation}

\begin{figure}
\includegraphics[clip=true,width=0.45\textwidth]{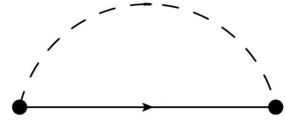}
\caption{Second order diagram for electron self -- energy.
Dashed line represents the Green's function of a quantum fluctuation $F$, 
continuous line represents the electronic Green's function $G$.}
\label{SE}
\end{figure}

Electron Green's function in Matsubara representation is written in a standerd
form:
\begin{equation}
G(\varepsilon_n,{\bf p})=\frac{1}{i\varepsilon_n-\xi_{\bf p}-
\Sigma(\varepsilon_n,{\bf p})}
\label{GrF}
\end{equation}
where $\varepsilon_n=(2n+1)\pi T$, $\xi_{\bf p}$ is the free -- electron
spectrum counted from the Fermi level (chemical potential), and self -- energy
part can be taken in the simplest approximation shown by diagram in Fig. \ref{SE}:
\begin{eqnarray}
&& \Sigma(\varepsilon_n,{\bf p})=-\frac{T}{N}\sum_{\bf q}g^2_{\bf q}\sum_{m}
F({\bf q},i\omega_m)
G(\varepsilon_n+\omega_m,{\bf p+q})\nonumber\\
&& =-\frac{T}{N}\sum_{\bf q}g^2_{\bf q}\sum_{m}\int_{-\infty}^{\infty}d\omega
\frac{S({\bf q},\omega)}
{i\omega_m-\omega}
\left(1-e^{-\frac{\omega}{T}}\right)\times\nonumber\\
&& \times\frac{1}{i\varepsilon_n
+i\omega_m-\xi_{\bf p+q}}
\label{SEFG}
\end{eqnarray}
where $g_{\bf q}$ is the coupling constant (matrix element of interaction
potential).

Consider now the case when the average frequency of fluctuation $<\Omega>$
is much lower than the temperature $T$, i.e. the classic limit for fluctuations.
Then in Eq. (\ref{SEFG}) it is sufficient to take into consideration only the
term with $m=0$, going actually to the picture of {\em quasi -- elastic} 
scattering by fluctuations:
\begin{eqnarray}
\Sigma(\varepsilon_n,{\bf p})=\frac{1}{N}\sum_{\bf q}g^2_{\bf q}\int_{-\infty}^{\infty}d\omega
\frac{S({\bf q},\omega)}
{i\varepsilon_n-\xi_{\bf p+q}}=\nonumber\\
=\sum_{\bf q}g^2_{\bf q}S({\bf q})\frac{1}
{i\varepsilon_n-\xi_{\bf p+q}}
\label{Sen}
\end{eqnarray}
where we have introduced the structure factor of fluctuations \cite{PinNoz}:
\begin{equation}
S({\bf q})=\frac{1}{N}\int_{-\infty}^{\infty}d\omega S({\bf q},\omega)
\label{StrFac}
\end{equation}
In fact, this is in a direct analogy with the well--known  Ziman -- Edwards
approximation in the theory of liquid metals \cite{Ziman,Edw}. The case of
$S({\bf q})=const$ corresponds to totally chaotic distribution of static
scattering centers \cite{Diagram}.

\subsubsection{Phonons}

In case of scattering by phonons fluctuation operator $\rho_{\bf q}$ can be
expressed via (Bose) creation and annihilation operators of phonons as
\cite{PinNoz}:
\begin{equation}
\rho_{\bf q}=\frac{1}{\sqrt 2}\left(b^+_{\bf q}+b_{-\bf q}\right)
\label{phons}
\end{equation}
Then:
\begin{eqnarray}
S({\bf q},\omega)=Z^{-1}\sum_m e^{-\frac{E_m}{T}}\left[<m|b_{\bf q}b^+_{\bf q}|m>
\delta(\omega-\omega_{\bf q})+\right.\nonumber\\
\left.+<m|b^+_{-\bf q}b_{-\bf q}|m>\delta(\omega+
\omega_{\bf q})\right]\nonumber\\
\label{Sqw}
\end{eqnarray}
where $\omega_{\bf q}$ is phonon spectrum.
Introducing the usual Bose distribution:
\begin{equation}
n_{\bf q}=Z^{-1}\sum_m e^{-\frac{E_m}{T}}<m|b^+_{\bf q}b_{\bf q}|m>=
\frac{1}{e^{\frac{\omega_{\bf q}}{T}}-1}
\label{bose}
\end{equation}
we get \cite{PinNoz}
\begin{eqnarray}
S({\bf q},\omega)=\left[(n_{\bf q}+1)\delta(\omega-\omega_{\bf q})
+n_{\bf q}\delta(\omega+\omega_{\bf q})\right]+\nonumber\\
+\delta(\omega-\omega_{\bf q})+n_{\bf q}\left[\delta(\omega-\omega_{\bf q})
+\delta(\omega+\omega_{\bf q})\right]
\label{Swqq}
\end{eqnarray}
Under the conditions of $T\gg\omega_{\bf q}$, we go to the classical limit
(equipartition theorem):
\begin{equation}
n_{\bf q}=\frac{T}{\omega_{\bf q}}
\label{qclass}
\end{equation} 
and accordingly
\begin{equation}
S({\bf q})=\frac{2T}{\omega_{\bf q}}
\end{equation}
Thus we obtain the structure factor linearly growing with temperature, while
its momentum dependence is determined by phonon spectrum.
Then:
\begin{equation}
\Sigma(\varepsilon_n,{\bf p})=T\sum_{\bf q} \frac{2 g_{\bf q}^2}{\omega_{\bf q}}
\frac{1}{i\varepsilon_n-\xi_{\bf p+q}}
\label{SEel}
\end{equation}
To simplify the model further, let us assume the the phonon spectrum is
dispersionless (Einstein phonon or optical phonon with very weak dispersion),
taking $\omega_{\bf q}=\Omega_0$. Then performing all calculations as in the problem
of an electron in a system of random impurities  \cite{Diagram}, we get:
\begin{equation}
\Sigma(\varepsilon_n,{\bf p})=-i\pi sign \varepsilon_n\frac{2g_0^2}{\Omega_0}N(0)T
\label{SiGG}
\end{equation}
where $N(0)$ is the density of states on the Fermi level.
Correspondingly, the damping is written as:
\begin{equation}
\frac{\Gamma(T)}{2}=\pi\frac{2g_0^2}{\Omega_0}N(0)T=\pi\lambda_0 T
\label{GamT}
\end{equation}
where we have introduced the usual dimensionless coupling constant of
electron -- phonon interaction:
\begin{equation}
\lambda_0=\frac{2g^2_0N(0)}{\Omega_0}
\label{Lamb_st}
\end{equation}
After standard calculations \cite{Diagram} we obtain the resistivity as:
\begin{equation}
\rho(T)=\frac{m}{ne^2}\Gamma(T)=2\pi\lambda_0\rho_{P}(T)
\label{ResistT}
\end{equation}
which is essentially the high -- temperature limit of Eliashberg -- McMillan
theory (\ref{RphMc}). Now the constant $\alpha$ used in the definition of
``Planckian'' relaxation time (\ref{Planck}) is expressed via the parameters
of the theory as:
\begin{equation}
\alpha=2\pi\lambda_0
\label{alP}
\end{equation}
Naturally it is not universal and just proportional to the coupling constant.

Practically the same result can be easily obtained in Eliashberg -- McMillan
approximation \cite{Diagram}, where the expression for Matsubara self -- energy
is written as:
\begin{eqnarray}
\Sigma(\varepsilon_n)=\int d\varepsilon'\int_{0}^{\infty} d\omega\alpha^2(\omega)
F(\omega)\times\nonumber\\
\times\Biggl\{\frac{f(\varepsilon') + n(\omega)}
{i\varepsilon_n - \varepsilon'+\omega}
+ \frac{1-f(\varepsilon') + n(\omega)}
{i\varepsilon_n - \varepsilon'-\omega}\Biggr\}
\label{self-energy_3}
\end{eqnarray}
where $f(\varepsilon)$ and $n(\omega)$ are standard Fermi and Bose distributions.
In the high -- temperature limit $T\gg\omega$ Eq. (\ref{self-energy_3}) reduces
to:
\begin{eqnarray}
&& \Sigma(\varepsilon_n)=\int d\varepsilon'\int_{0}^{\infty} d\omega\alpha^2(\omega)
F(\omega)\times\nonumber\\
&& \times\frac{T}{\omega}\Biggl\{\frac{1}
{i\varepsilon_n - \varepsilon'+\omega}
+ \frac{1}
{i\varepsilon_n - \varepsilon'-\omega}\Biggr\}=\nonumber\\
&& =T\int_{-\infty}^{\infty}d\varepsilon'2\int_{0}^{\infty}\frac{d\omega}{\omega}
\alpha^2(\omega)F(\omega)\frac{1}{i\varepsilon_n-\varepsilon'}=\nonumber\\
&& =-i\pi\lambda T sign\varepsilon_n
\label{SEMcMill}
\end{eqnarray}
where we have introduced the standard definition of the coupling constant
of Eliashberg -- McMillan theory (\ref{lamb_McM}), thus reproducing the result
like (\ref{GamT}). For resistivity we get again Eq. (\ref{ResistT}) with an
obvious replacement $\lambda_0\to\lambda$, where $\lambda$ is defined in
(\ref{lamb_McM}).\footnote{Here and in the following we neglect for brevity the
insignificant for our aims difference between $\lambda$ and $\lambda_{tr}$.}

\subsubsection{More general model}

Now let us try to avoid explicit introduction of phonons (or any other quasiparticles
related to fluctuations). From Eq. (\ref{AS}) for $\omega\ll T$ we get:
\begin{equation}
A({\bf q},\omega)\approx \frac{\omega}{T}S({\bf q},\omega)
\label{Aq}
\end{equation}
or
\begin{equation}
S({\bf q},\omega)\approx \frac{T}{\omega}A({\bf q},\omega)
\label{Sq}
\end{equation}
Substituting this expression to Eq. (\ref{Sen}) we obtain the following
expression for self -- energy:
\begin{equation}
\Sigma(\varepsilon_n,{\bf p})=\frac{T}{N}\sum_{\bf q}g^2_{\bf q}
\int_{-\infty}^{\infty}\frac{d\omega}{\omega}
\frac{A({\bf q},\omega)}
{i\varepsilon_n-\xi_{\bf p+q}}
\label{Senr}
\end{equation}
where everything is determined by the spectral density of fluctuations
$A({\bf q},\omega)$, which does not necessarily of quasiparticle form.
Naturally, for the simplest model with $A({\bf q},\omega)=\delta(\omega-\Omega_0)$ 
(Einstein model for fluctuations) from (\ref{Senr}) immediately follow the
Eqs. (\ref{SiGG}) -- (\ref{Lamb_st}) derived above. 
In case of no $\bf q$ dependence from (\ref{Senr}) we immediately obtain:
\begin{eqnarray}
&& \Sigma(\varepsilon_n,{\bf p})=\frac{T}{N}\sum_{\bf q}g^2_0\int_{-\infty}^{\infty}
\frac{d\omega}{\omega}
\frac{A(\omega)}
{i\varepsilon_n-\xi_{\bf p+q}}=\nonumber\\
&& =-i\pi sign\varepsilon_n N(0)\frac{T}{N}g^2_{0}\int_{-\infty}^{\infty}
\frac{d\omega}{\omega}
A(\omega)=\nonumber\\
&& =-i\pi sign \varepsilon_n\frac{2g_0^2}{<\Omega>}N(0)T
\label{SiGGG}
\end{eqnarray}
where we have introduced an average frequency of fluctuations:
\begin{equation}
<\Omega>^{-1}=2\int_{0}^{\infty}\frac{d\omega}{\omega}A(\omega)
\label{avFr}
\end{equation}
This result is actually equivalent to (\ref{SiGG}).

In general case, when we can not neglect the momentum dependence of spectral
density of fluctuations, we can use Eliashberg -- McMillan approach, assuming
that fluctuations scatter electrons in some narrow ($\sim T$) layer around
the Fermi surface. Then we can introduce the self -- energy averaged over the
momenta {\em on the Fermi surface}: 
\begin{equation}
\Sigma(\varepsilon_n)=\frac{1}{N(0)}\sum_{\bf p}\delta{(\xi_{\bf p})}
\Sigma(\varepsilon_n,{\bf p}),
\label{SigMac}
\end{equation} 
and also an effective (averaged over initial and final momenta on the Fermi
surface) interaction:
\begin{eqnarray}
&& g_{\bf pp'}^2A({\bf p-p',\omega})\Longrightarrow
\nonumber\\
&& \frac{1}{N(0)}\sum_{\bf p}\frac{1}{N(0)}\sum_{\bf p'}g_{\bf pp'}^2
A({\bf p-p',\omega})
\delta(\xi_{\bf p})\delta({\xi_{\bf p'}})\nonumber\\
&& \equiv\frac{1}{N(0)}\alpha^2(\omega)F(\omega)
\label{McMEavr}
\end{eqnarray}
where
\begin{equation}
F(\omega)=\sum_{\bf q}A({\bf q,\omega})
\label{flDOS}
\end{equation}
is the density of states of fluctuations. Then from (\ref{Senr}) we obtain 
for (\ref{SigMac}):
\begin{eqnarray}
&& \Sigma(\varepsilon_n)=\frac{T}{N(0)}\int_{-\infty}^{\infty}
\frac{d\omega}{\omega}\alpha^2(\omega)F(\omega)N(0)\int_{-\infty}^{\infty} d\xi\frac{1}
{i\varepsilon_n-\xi}=\nonumber\\
&& =-i\pi sign\varepsilon_n T\int_{-\infty}^{\infty}\frac{d\omega}{\omega}
\alpha^2(\omega)F(\omega)=\nonumber\\
&& =-i\pi sign\varepsilon_n\lambda T\equiv -i\frac{\Gamma(T)}{2}sign\varepsilon_n
\label{SE-MCM}
\end{eqnarray}
where we again introduced the dimensionless coupling constant as in Eliashberg --
McMillan theory defined by Eq. (\ref{lamb_McM}), which value is in fact
determined by the (averaged as in (\ref{McMEavr})) spectral density of
fluctuations $A({\bf q},\omega)$, which does not necessarily describes any 
quasiparticles. 

Finally we obtain:
\begin{equation}
\Gamma(T)=2\pi\lambda T
\label{Plkrel}
\end{equation}
which is of the same form as Eq. (\ref{GamT}) and leads immediately to
(\ref{ResistT}).
Strictly speaking this is not necessarily so if we remember the possible
temperature dependence of spectral density $A({\bf q},\omega)$. Planckian
distribution is obtained only in the absence or weakness of this dependence.
It is obvious that Eq. (\ref{GamT}) immediately follows from  (\ref{SE-MCM}) 
and (\ref{lamb_McM}) if $F(\omega)=\delta(\omega-\Omega_0)$, which corresponds
to Einstein spectrum of fluctuations.

\subsubsection{Quantum critical point and around}

The simplest model considered above certainly does not explain the $T$ -- linear
behavior of resistivity in cuprates and similar systems in normal state,
starting from very low temperatures. It should be noted, that this behavior is
often related to the closeness of these systems to some quantum critical point.
Consider the schematic phase diagram of hole -- doped cuprates shown in Fig.
\ref{P_Diagr} \cite{Tail}.
\begin{figure}
\includegraphics[clip=true,width=0.45\textwidth]{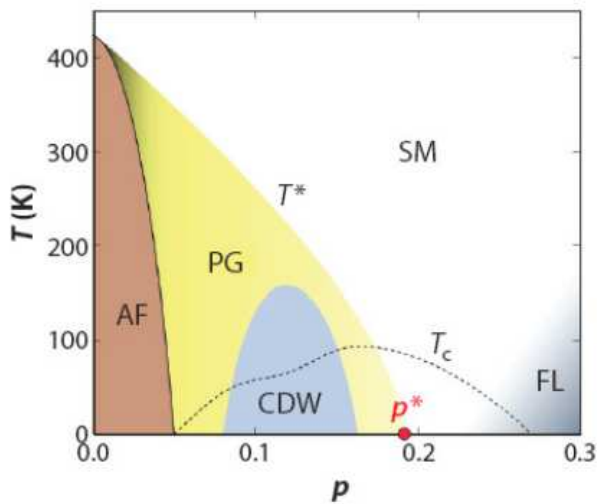}
\caption{Typical phase diagram of hole -- doped cuprates. Superconducting state
is observed in the region of temperatures below $T_c$, shown by dashed line.
Antiferromagnetic region AF is shown in brown. Pseudogap phase PG exists at
temperatures below $T^{\star}$ (yellow) and ends at $T=0$ in the critical
point $p^{\star}$ (red dot). Inside the pseudogap phase exists the phase with
charge density wave CDW (blue). The region of strange metal SM (white)
continuously transforms into the Fermi -- liquid region FL at higher dopings.
\cite{Tail}}
\label{P_Diagr}
\end{figure}
\begin{figure}
\includegraphics[clip=true,width=0.45\textwidth]{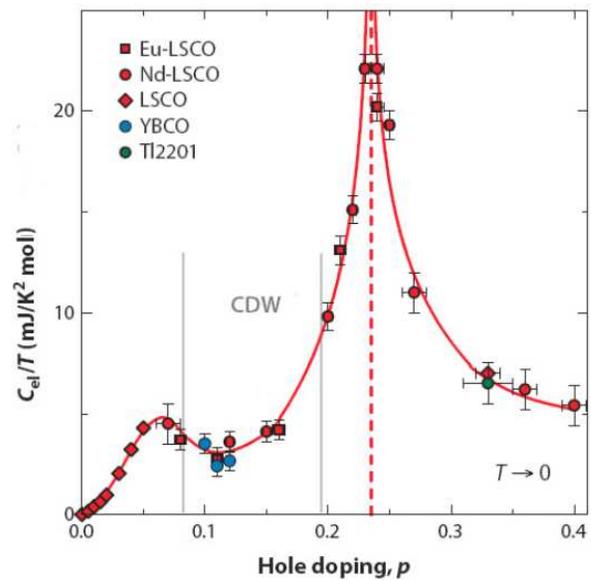}
\caption{Electronic specific heat coefficient 
$\gamma=C_{el}/T$ 
in normal state ($T=$5K) in the vicinity of the quantum critical point
(superconductivity is suppressed by external magnetic field) as function of
doping $p$ for La$_{1.8-x}$Eu$_{0.2}$Sr$_x$CuO$_4$ (Eu-LSCO, red squares),
La$_{1.6-x}$Nd$_{0.4}$Sr$_x$CuO$_4$ (Nd-LSCO, red circles),
La$_{2-x}$Sr$_x$CuO$_4$ (LSCO, red diamonds), YBa$_2$Cu$_3$O$_y$
(YBCO, blue circles) and Tl$_2$Ba$_2$CuO$_{6-\delta}$ (Tl2201, green circles)
\cite{Tail}}
\label{C_qcrp}
\end{figure}
This diagram is well -- known for a long time, but what is important for us now
is the possible existence here of a quantum critical point
\cite{Sachd,Stish} at carriers (holes) concentration $p=p^{\star}$, where
the pseudogap region ends \cite{SadPG,SadLPI,GDMFT}.
Experimental evidence in favor of the existence of such a critical point are
rather numerous \cite{Tail}. As an example in Fig. \ref{C_qcrp} we just show one of the 
most striking -- the sharp singularity of the electronic specific heat 
coefficient $\gamma=C_{el}/T$ at $p=p^{\star}$ in the normal state of cuprates,
obtained in a strong magnetic field suppressing superconductivity \cite{Tail}.

The nature of the pseudogap state in cuprtaes is not completely clarified by now.
In particular, it is not clear, whether PG region on the phase diagram of
Fig. \ref{P_Diagr} is some new phase and the line $T^{\star}$ defines the
critical temperature of a true phase transition, or it is the region of some
crossover to antiferromagnetic phase AF, with well developed fluctuations of
AF short -- range order \cite{SadPG,SadLPI,GDMFT}. antiferromagnetic scenario
of pseudogap formation also has a serious experimental support \cite{Tail}, 
though some authors believe, that the transition to pseudogap phase is some
true phase transition with some still unknown order parameter.
A number of specific models  (see e.g. \cite{Varma}) of such phase transition
were proposed, allowing to consider $p=p^{\star}$ as a true quantum critical
point. However, it is still unclear, whether we can speak of the quantum
critical point in crossover scenario. In any case, the $T$ -- linear behavior 
of resistivity in cuprates is observed close to optimal doping, which nearly
coincides with $p=p^{\star}$. To the left of this concentration resistivity
demonstrates some dielectric (localization) effects (the strong negative
temperature coefficient of resistivity) \cite{SadLoc}, while to the right
we have a crossover to more or less usual Fermi -- liquid -- like behavior
with quadratic growth of resistivity with temperature.

\begin{figure}
\includegraphics[clip=true,width=0.45\textwidth]{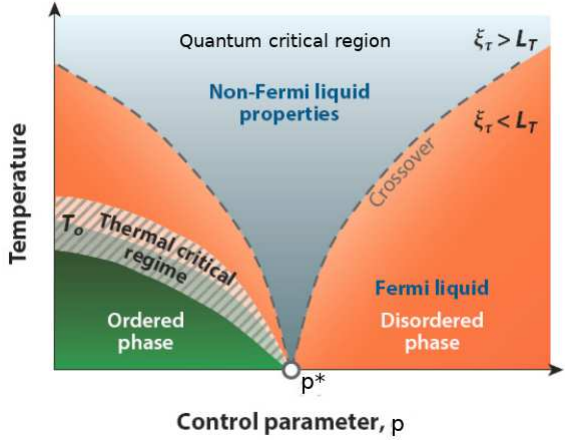}
\caption{Phase diagram in the vicinity of the quantum critical point.
Non -- Fermi -- liquid behavior of relaxation rate and resistivity can be
observed in the quantum critical region, where 
$\xi_{\tau}>L_T=\frac{\hbar}{k_BT}$, which corresponds to $\hbar\Omega_0<k_BT$}.
\label{P_QCP}
\end{figure}

the closeness of the system to the quantum critical point can, in principle,
explain the $T$ -- linear growth of resistivity even within the elementary
model considered above, To observe such growth at low temperatures, it is
sufficient to satisfy the inequality $\Omega_0\ll T$, where $\Omega_0$ is a
characteristic frequency of fluctuations, scattering the electrons.
In the vicinity of the quantum critical point (of any nature) we can expect
the typical ``softening'' of the appropriate fluctuation mode according to the
relation \cite{Sachd,Stish}:
\begin{equation}
\Omega_0\sim |p-p^{\star}|^{z\nu}
\label{softm}
\end{equation}
where $p$ is e.g. the concentration of carriers (holes of electrons).
Here $\nu$ and $z$ are the standard critical exponents of the theory of quantum
phase transitions, determining the critical behavior of characteristic lengths:
\begin{equation}
\xi\sim|p-p^{\star}|^{-\nu},\ \xi_{\tau}\sim|p-p^{\star}|^{-z\nu}
\label{lengths}
\end{equation}
Here $\tau$ denotes the imaginary (Matsubara) time and  above we just
put $\Omega_0\sim\xi_{\tau}^{-1}$.

As shown in Fig. \ref{P_QCP} the quantum critical region is defined by 
inequality \cite{Sachd,Stish}:
\begin{equation}
\xi_{\tau}>\frac{1}{T} \quad\mbox{or}\quad \Omega_0<T
\label{QcrP}
\end{equation}
which may guarantee the $T$ -- linear behavior of resistivity in this region in
cuprates or similar systems. However, the nature of quantum fluctuations of
importance here, as well as the mechanism of their interaction with electrons
remain an open question.

\section{General relations for the Green's function and conductivity}

\subsubsection{Green's function fir the system of interacting particles}

Let us remind some general expressions for an arbitrary system of interacting
electrons. We have the following general expression for Matsubara Green's
function \cite{AGD}:
\begin{equation}
G(i\varepsilon_n,{\bf p})=\frac{1}{i\varepsilon_n-\xi_{\bf p}-
\Sigma(i\varepsilon_n,{\bf p})}
\label{A1}
\end{equation}
Having in mind the electron -- phonon coupling and its analogs
(quantum fluctuations of a general nature), dropping the momentum dependence
of self -- energy, we can write down the following expression in the standard
notations of Eliashberg -- McMillan theory: 
\begin{equation}
\Sigma(i\varepsilon_n)=[1-z(\varepsilon_n)]i\varepsilon_n+\chi(\varepsilon_n)
\label{A15}
\end{equation}
Then:
\begin{eqnarray}
G(i\varepsilon_n,{\bf p})=\frac{1}{z(\varepsilon_n)i\varepsilon_n-\xi_{\bf p}-
\chi(\varepsilon_n)}=\nonumber\\
=\frac{Z(\varepsilon_n)}{i\varepsilon_n-Z(\varepsilon_n)\xi_{\bf p}-Z(\varepsilon_n)
\chi(\varepsilon_n)}
\label{A16}
\end{eqnarray}
where we have defined:
\begin{equation}
Z(\varepsilon_n)=z^{-1}(\varepsilon_n)
\label{A17}
\end{equation}
Let us define also:
\begin{equation}
\varepsilon({\bf p},\varepsilon_n)=Z(\varepsilon_n)\xi_{\bf p}+Re\chi(\varepsilon_n)
\label{A18}
\end{equation}
\begin{equation}
\Gamma_{i}(T)sign\varepsilon_n=-2Z(\varepsilon_{n})Im\chi(\varepsilon_n)
\label{A19}
\end{equation}
Then the Green's function can be written in the following general enough form:
\begin{equation}
G(i\varepsilon_n,{\bf p})=\frac{Z(\varepsilon_n)}{i\varepsilon_n-\varepsilon({\bf p},
\varepsilon_n)+\frac{i}{2}\Gamma(T)sign\varepsilon_n}
\label{A20}
\end{equation}
If there is no dependence on $\varepsilon_n$ in (\ref{A18}) we can write the
Green' function in the usual form:
\begin{equation}
G(i\varepsilon_n,{\bf p})=\frac{Z}{i\varepsilon_n-\varepsilon({\bf p})
+\frac{i}{2}\Gamma(T)sign\varepsilon_n}
\label{A21}
\end{equation}
which is sufficient for our aims.

These expressions more or less correspond to the usual picture of electron --
phonon interaction, when $\lambda\ll 1$, though we can try to ``drag'' them
to the region of $\lambda>1$, using the Migdal theorem \cite{Diagram}, 
which allows to neglect vertex corrections. In particular, in this approach 
the renormalization factor is: \cite{Diagram}
\begin{equation}
Z=\frac{1}{1+\lambda}
\label{A22}
\end{equation}
and does not depend explicitly on $\varepsilon_n$. Eq. (\ref{A22}) is valid
within an energy layer of the order of double Debye frequency around the
Fermi level (in case of interaction with phonons) or the double average 
frequency of fluctuations $<\Omega>$. Outside this layer it is obvious
that $Z=1$ -- fluctuations are practically irrelevant for electrons of 
high -- enough energy.

\subsubsection{General expressions for conductivity}

Diagonal element of conductivity tensor at ${\bf q}=0$ can be written
as: \cite{Diagram,Fuku}
\begin{equation}
\sigma_{xx}(\omega)=\frac{1}{i\omega}\left\{
\Phi_{xx}(\omega+i\delta)-\Phi_{xx}(0+i\delta)\right\}
\label{condxx}
\end{equation}
where:
\begin{eqnarray}
&& \Phi_{xx}(i\omega_m)=-2eT\sum_n\sum_{\bf p}\frac{p_x}{m}J_x({\bf p,p,}
\varepsilon_n,\varepsilon_n+\omega_m)\times\nonumber\\
&& \times G(\varepsilon_n{\bf p})G(\varepsilon_n+\omega_m{\bf p})
\label{Phixx}
\end{eqnarray}
which corresponds to the usual loop diagram shown in Fig. \ref{Jloop}, where
the ``current'' vertex can be written as:
\begin{equation}
J_{\mu}({\bf p,p},\varepsilon_n,\varepsilon_n+\omega_m)\equiv
\frac{e}{m}p_{\mu}\Xi({\bf p},\varepsilon_n,\varepsilon_n+\omega_m)
\label{Jcurr}
\end{equation}
and the ``bare'' current vertex is $\frac{e}{m}p_{\mu}$.
\begin{figure}
\includegraphics[clip=true,width=0.45\textwidth]{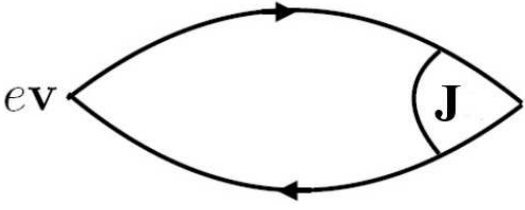}
\caption{Loop diagram determining conductivity.}
\label{Jloop}
\end{figure}

Then rewriting Eq. (\ref{Phixx}) as:
\begin{eqnarray}
&& \Phi_{xx}(i\omega_m)=-2e^2T\sum_n\sum_{\bf p}\frac{p_x^2}{m^2}
\Xi({\bf p},\varepsilon_n,\varepsilon_n+\omega_m)\times\nonumber\\
&& \times G(\varepsilon_n{\bf p})G(\varepsilon_n+\omega_m{\bf p})
\label{Phixxx}
\end{eqnarray}
and performing the standard summation over $n$ and analytic continuation
$i\omega_m\to\omega+i\delta$ \cite{Fuku}, we obtain the static conductivity
at $(\omega\to 0)$:
\begin{equation}
\sigma_{xx}=\frac{e^2}{2\pi}\sum_{\bf p}\frac{p_x^2}{m^2}\Xi({\bf p})
G^R({\bf p},0)G^{A}({\bf p},0)
\label{sigmstatic}
\end{equation}
where we have introduced the static limit 
$\Xi({\bf p})=\Xi({\bf p},0-i\delta,0+i\delta)$.

Obviously, the main difficulty here is related to the explicit calculation of
the vertex part for the system of interacting particles. The simplest estimate
can be obtained using in Eq. (\ref{Phixx}) the obvious Ward identity
\cite{Fuku}, which is valid for $\omega_m=0$:
\begin{eqnarray}
&& J_{\mu}({\bf p,p},\varepsilon_n,\varepsilon_n)=
-e\frac{\partial}{\partial p_{\mu}}G^{-1}(\varepsilon_n,{\bf p})=\nonumber\\
&& =-e\left[-v_{\bf p}^{\mu}-\frac{\partial\Sigma(\varepsilon_n,{\bf p})}
{\partial p_{\mu}}\right]
\label{JWard}
\end{eqnarray}
where velocity $v_{\bf p}^{\mu}=\frac{\partial\xi_{\bf p}}{\partial p_{\mu}}$.
If we assume here that the self -- energy is momentum independent i.e.
$\Sigma(\varepsilon_n,{\bf p})=\Sigma(\varepsilon)$, as it usually takes
place for electron -- phonon interaction, the vertex (\ref{JWard}) is reduced
to the ``bare'' one:
\begin{equation}
J_{\mu}({\bf p,p},\varepsilon_n,\varepsilon_n)=
ev_{\bf p}^{\mu}=\frac{e}{m}p_{\mu}
\label{JWard0}
\end{equation}
which corresponds to $\Xi({\bf p},\varepsilon_n,\varepsilon_n)=1$.
Using (\ref{JWard}) in (\ref{Phixx}) we get:
\begin{eqnarray}
&& \Phi_{xx}(i\omega_m)=
-2e^2T\sum_n\sum_{\bf p}\frac{p_x}{m}\left[\frac{\partial}{\partial p_{\mu}}
G^{-1}(\varepsilon_n,{\bf p})\right]\times\nonumber\\
&& \times G(\varepsilon_n{\bf p})G(\varepsilon_n+
\omega_m{\bf p})=
\nonumber\\
&& =-2e^2T\sum_n\sum_{\bf p}\frac{p_x^2}{m^2}G(\varepsilon_n{\bf p})
G(\varepsilon_n+\omega_m{\bf p})
\label{PhixxxW}
\end{eqnarray}
and then, making all transformations as in going from  (\ref{Phixx}) to 
(\ref{sigmstatic}) we obtain the following expression approximate relation
for static conductivity:
\begin{equation}
\sigma_{xx}=\frac{e^2}{2\pi}\sum_{\bf p}\frac{p_x^2}{m^2}
G^R({\bf p},0)G^{A}({\bf p},0)
\label{sigmstatic0}
\end{equation}
which simply corresponds to the choice of $\Xi({\bf p})=1$ in (\ref{sigmstatic}). 
In contrast to (\ref{sigmstatic}) this is  certainly a kind of  approximation.
It is used e.g. in Ref. \cite{Varma}. What is lost here will become clear
below.

Further calculations will be done for the general case of (\ref{sigmstatic}). 
Keeping in mind the typical metal, where all the physics of conductivity is
determined in the vicinity of the Fermi surface (and scattering is rather weak),
we can use the usual integration over the energy spectrum linearized close
to the Fermi level and write (\ref{sigmstatic}) as ($Re\Sigma(0)$ can be
included in the renormalization of the chemical potential):
\begin{eqnarray}
&& \sigma_{xx}=\frac{e^2}{2\pi}N(0)\int_{-\infty}^{\infty}d\xi_{p}\frac{1}{3}
\frac{p_F^2}{m^2}\Xi(p_F)G^R({\bf p},0)G^{A}({\bf p},0)=\nonumber\\
&& =\frac{e^2}{2\pi}\frac{1}{3}v_F^2\Xi(p_F)N(0)\int_{-\infty}^{\infty}d\xi_{p}
\frac{1}{\xi_p^2+Im^2\Sigma(0)}=\nonumber\\
&& =e^2\frac{1}{3}v_F^2N(0)\frac{\Xi(p_F)}{2Im\Sigma(0)}
\equiv\frac{ne^2}{m}\tau_{tr}
\label{sigmstat}
\end{eqnarray}
From this expression it is clear, by the way, that approximation
$\Xi(p_F)=1$ corresponds to neglecting the difference between $\tau$ and 
$\tau_{tr}$, which we used everywhere above. For the case of scattering by
point -- like impurities or Einstein phonons this is simply always valid --
from (\ref{sigmstat}) we immediately get (\ref{ResistT}). 
Precisely the same expression for conductivity is used e.g. in Ref. \cite{Varma}. 
Now it is clear that the mass renormalization, appearing e.g. in the expression
for electronic specific heat {\em does not enter} the expression for 
$\sigma_{xx}$, where everything is determined by the ``bare'' (band) mass.
This result is well -- known actually for a long time \cite{Naka,PrKad,Hein,Grim}.

\footnotesize
Contribution from the possible momentum dependence of 
$\Sigma(\varepsilon,{\bf p})$ into static conductivity also can be obtained
in rather simple way \cite{Fuku}. Let us write:
\begin{equation}
G^R(\varepsilon,{\bf p})=[Z^{-1}\varepsilon-b^{-1}\xi_{p}
+Im\Sigma(\varepsilon,p_F)]^{-1}
\label{GR}
\end{equation}
where as usual we introduced:
\begin{equation}
Z^{-1}=1-\frac{\partial}{\partial\varepsilon}
Re\Sigma(\varepsilon,p_F)|_{\varepsilon=0}
\label{Z1}
\end{equation}
\begin{equation}
b^{-1}=1+\frac{\partial}{\partial\xi_p}Re\Sigma(0,{\bf p})|_{|\bf p|=p_F}
\label{b1}
\end{equation}
Then, using Eq. (\ref{GR}), we obtain density of stets renormalization on
the Fermi level and get:
\begin{equation}
\tilde N(0)=-\frac{1}{\pi}\sum_{\bf p}Im G^R(0,{\bf p})
=\frac{1}{\pi}\int_{-\infty}^{\infty}d\xi_p\frac{1}{b^{-2}\xi_p+
Im\Sigma(0,{\bf p})}
=bN(0)
\label{DOSr}
\end{equation}
Now we may repeat all calculations for conductivity done above using (\ref{GR}), 
and obtain the final result for static conductivity as: \cite{Fuku}:
\begin{equation}
\sigma_{xx}=\frac{ne^2}{m}\frac{\Xi(p_F)}{2Im\Sigma(0)}b=\frac{ne^2}{m}
\tau_{tr}\frac{\tilde N(0)}{N(0)}
\label{sigmstatb}
\end{equation}

\normalsize

\subsubsection{Self -- consistent calculation}

Let us return to the simplest model of electron -- phonon interaction and
make the self -- consistent calculation, when electron line on diagram of
Fig. \ref{SE} is ``dressed'', i.e. takes into account all orders of non --
intersecting interaction lines. Again we assume the validity of Migdal's
theorem \cite{Diagram}. Acting in the spirit of Eqs. (\ref{A15}) -- (\ref{A21}), 
we can write the expression for the Green's function as (simplified variant of 
Eq. (\ref{A20})):
\begin{equation}
G(\varepsilon_n,{\bf p})=\frac{Z}{i\varepsilon_n-Z\xi_{\bf p}
+\frac{i}{2}Z\Gamma(T)sign \varepsilon_n}
\label{G_dressed}
\end{equation}
which corresponds to a choice of $Z(\varepsilon_n)=Z$ 
$Re \chi(\varepsilon_n)=0$, $Im \chi(\varepsilon_n)=-\Gamma(T)$ in 
Eq. (\ref{A16}). Renormalization factor $Z<1$ is assumed to be a constant
for simplicity. Then it may seem, that taking $\Gamma_{i}(T)=2\pi\lambda T$ 
we may introduce the renormalized damping as:
\begin{equation}
\tilde\Gamma(T)=Z\Gamma(T)=2\pi\lambda ZT=2\pi\frac{\lambda}{1+\lambda}T
\label{tildGam}
\end{equation}
where we have accounted for the known result due  electron -- phonon
interactions in {\em low -- temperature} limit \cite{Diagram}:
\begin{equation}
Z=\frac{1}{1+\lambda}
\label{Zphon}
\end{equation}
Accordingly, for $\lambda\ll 1$ we have (\ref{GamT}), and for $\lambda\gg 1$
we get:
\begin{equation}
\tilde\Gamma(T)=2\pi T
\label{PlanckGam}
\end{equation}
i.e. the {\em universal} ``Planckian'' behavior of relaxation rate (\ref{Planck}) 
with $\alpha=2\pi$ {\em independent} of coupling constant of electrons with
fluctuations (phonons).  In general case, for arbitrary values of $\lambda$ 
we have $\alpha<2\pi$, so that the upper limit for $\alpha$ appears in a
natural way and is defined by (\ref{PlanckGam}).

However, substituting  (\ref{G_dressed}) into (\ref{SEel}), which is obtained
in  {\em high -- temperature} limit, in the model with Einstein spectrum of
fluctuations (phonons) we immediately obtain:
\begin{equation}
\Sigma(\varepsilon_n,{\bf p})=\Sigma(\varepsilon_n)=
-i\pi \frac{2g_0^2}{\Omega_0}N(0)Tsign \varepsilon_n=
-\frac{i}{2}\Gamma(T)sign \varepsilon_n
\label{SEZGam}
\end{equation}
so that the renormaliztion factor $Z$ in damping is {\em canceled out} 
and we just reproduce the usual result derived above without any renormalizations.
This is not surprising at all --- it is clear from the very beginning that
$Z\approx 1$ for temperatures (energies) much higher than characteristic phonon
frequencies (or any other fluctuation quanta scattering the electrons).
Similar results are valid not only in Einstein model, but also in the general
case, described by Eliashberg -- McMillan approximation
(\ref{SEMcMill}), (\ref{SE-MCM}) \cite{Sad20}.

It may also seem, that limitations discussed above may be obtained from
completely different considerations. Let us write down the general enough
expression for Matsubara Green's function in high -- temperature limit as:
\begin{equation}
G(i\varepsilon_n,{\bf p})=\frac{1}{i\varepsilon_n-\varepsilon_{\bf p}+\frac{i}{2}
\Gamma(T)sign\varepsilon_n}
\label{MatGr}
\end{equation}
After substitution here the limiting value of $\Gamma(T)$ from (\ref{Planck})
we obtain:
\begin{equation}
G(i\varepsilon_n,{\bf p})=\frac{1}{i\varepsilon_n-\varepsilon_{\bf p}+\frac{i}{2}
\alpha T sign\varepsilon_n}
\label{MatGrP}
\end{equation}
Then we immediately see that the constant $\alpha$, seemingly, can not achieve
the value of $\alpha=2\pi$, as Matsubara frequencies in (\ref{MatGrP}) become
in this case {\em even}, i.e. the system of fermions  ``turns'' into bosons,
which is just impossible because of the general spin -- statistics theorem ---
no interaction (leading e.g. to temperature -- dependent relaxation) can not
change the statistics of particles. Then, apparently, we have to introduce
the same limitation again:
\begin{equation}
\Gamma(T)=\frac{1}{\tau(T)}<2\pi T
\label{AlpPlanck}
\end{equation} 
that in particular means that in our model according to Eq. (\ref{Plkrel}) 
we always have an inequality $\lambda < 1$. Such limitation on coupling
constant agrees with recent results of quantum Monte -- Carlo calculations of 
electron -- phonon interactions \cite{Scal}, but contradicts many years of
experience in studies of superconductivity within Eliashberg -- McMillan
theory \cite{Savr,Alle}. Discussion of these contradictions and possible solutions
can be found in the recent paper \cite{ChuKiv}. 

In fact, the arguments given above are incorrect and the value of $\alpha$ can
can take any values including integers, with no paradoxes like the change of
statistics. These values are not special and the system may continuously pass
through them with the growth of $\alpha$ (i.e. the coupling strength). 
This can be easily understood making e.g. explicit  calculations of distribution
function (cf. Appendix).

It should be noted of course, that the $T$ -- linear behavior of damping in the
Green's function in all cases breaks the standard criteria for Fermi -- liquid
behavior \cite{PinNoz}, so that quasiparticles in the system are badly 
defined. This is also seen from the explicit form distribution function, which
is rather far from the usual Fermi step -- like function.


\subsubsection{Planckian relaxation delusion}

In Refs. \cite{Bruin,Legros} experimental data on resistivity were represented
by Drude expression (\ref{sigmainv}), where the effective mass $m^{\star}$ 
was determined from measurements of specific heat or de Haas -- van Alfen
effect, which in the model with electron -- phonon  or in a more general
model of scattering by quantum fluctuations of an arbitrary nature, is obtained
from the band structure effective mass by a simple substitution
$m\to m^{\star}=m(1+\lambda)$, which takes into account mass renormalization by
interactions (at low temperatures!). Inconsistency of this approach was already
stressed in Ref. \cite{Varma}. Leta us show, that precisely this representation
of experimental data leads to to the {\em delusion} of universal Planckian
relaxation in metals. In fact Eq. (\ref{ResistT}) for the 
{\em high -- temperature} limit of resistivity can be identically rewritten as:
\begin{equation}
\rho(T)=\frac{m(1+\lambda)}{ne^2}\frac{\Gamma(T)}{1+\lambda}=\frac{m^{\star}}
{ne^2}\tilde\Gamma(T)
\label{ResisT}
\end{equation}
where
\begin{equation}
\tilde\Gamma(T)=2\pi\frac{\lambda}{1+\lambda}T
\label{tildaGamma}
\end{equation}
which always leads to:
\begin{equation}
\tilde\Gamma(T)<2\pi T
\label{PlanckGamma}
\end{equation}
just imitating the universal ``Planckian'' behavior of relaxation rate
(\ref{Planck}) with $\alpha=2\pi$ as an upper limit independent of coupling
constant of electron with fluctuations (phonons).
The substitution of $m\to m^{\star}=m(1+\lambda)$ in (\ref{ResisT}) by itself
is correct, despite being used in an expression in high -- temperature limit.
Also correct is the treatment of experimental data in Refs. \cite{Bruin,Legros}, 
where they used the effective mass $m^{\star}$, obtained from 
{\em low -- temperature} measurements. However, this approach clouds the
crux of the matter, creating the delusion of universal Planckian behavior.

It is easy to estimate that the experimentally observed \cite{Bruin,Legros} 
values  of $\alpha\sim 1$ correspond to rather typical 
$\lambda\sim 0.2$, while $\alpha\sim 2.8$ for lead \cite{Bruin} corresponds
to $\lambda\sim 0.53$. Calculations within Eliashberg -- McMillan theory for
Pb give $\lambda\sim 1.5$ \cite{Savr}. For Nb in Ref. \cite{Bruin} was found
that $\alpha\sim 2.3$, which according to expressions given above corresponds to
$\lambda\sim 0.43$,  while the calculations \cite{Savr} give $\lambda\sim 1.26$. 
Suppression by a factor of $\sim 3$ of our values of $\lambda$ may be possibly
related to the fact, that in expressions for resistivity we must use
$\lambda_{tr}<\lambda$. However, the calculated values of $\lambda_{tr}$ 
\cite{Savr} equal 1.19 for Pb and 1.17 for Nb, which does not improve much 
the agreement with experiments.

Much more important may be the account of mass renormalization due to
electron -- electron interactions, which also contribute to electronic 
specific heat. which in practice is difficult to separate from phonon
contribution. Accordingly, Eq. (\ref{tildaGamma}) should be rewritten as:
\begin{equation}
\tilde\Gamma(T)=2\pi\frac{\lambda}{1+\lambda+\lambda_{ee}}T
\label{tildaGmm}
\end{equation}
where we have introduced $\lambda_{ee}$ -- the dimensionless parameter,
determining mass renormalization due to electron -- electron interaction.
In Landau -- Silin Fermi -- liquid theory $\lambda_{ee}= \frac{F^s_1}{3}$, 
where $F^s_1$ is the appropriate coefficient in the expansion of Landau
function \cite{PinNoz}. For typical metals $\lambda_{ee}\sim 1$. Then:
\begin{equation}
\alpha=\frac{2\pi\lambda}{1+\lambda+\lambda_{ee}}
\label{alpgener}
\end{equation}
so that taking as typical $\lambda\sim 0.4$ and $\lambda_{ee}\sim 1$ we 
immediately obtain $\alpha\sim 1.04$, while for $\lambda\sim 0.5$ we have
$\alpha\sim 1.25$, in goof agreement with majority of the data of 
Refs. \cite{Bruin,Legros}. For Pb, taking  $\lambda=1.68$ \cite{Savr} and 
$\lambda_{ee}=1$ we get $\alpha\sim 2.86$ in reasonable agreement with
``experimental'' value of $\alpha=2.8$ \cite{Bruin}. Similarly, for Nb we have
$\lambda=1.26$ \cite{Savr}, so that again using $\lambda_{ee}\sim 1$ we obtain
$\alpha\sim 2.42$, in good agreement with ``experimental'' value of $\alpha=2.3$ 
\cite{Bruin}. The interval of the values of  $\alpha=0.7-2.8$ \cite{Bruin,Legros} 
for $\lambda_{ee}=1$ corresponds to the interval of $\lambda=0.25-1.5$, that 
seems quite reasonable.

More detailed results of such estimates (assuming $\lambda_{ee}=1$) are given
in the Table:

\begin{tabular}{cccccc}
\hline

Metal & $\lambda$ & $\frac{2\pi\lambda}{1+\lambda}$ &
$\frac{2\pi}{1+\lambda+\lambda_{ee}}$ &  $\alpha_{exp}$ \\
\hline
Pb    &  1.68  &  3.93  &  2.86  &  2.8 \\
Nb    &  1.26  &  3.50  &  2.42  &  2.3 \\
Cu    &  0.14  &  0.77  &  0.41  &  1.0 \\
Al    &  0.44  &  1.91  &  1.13  &  1.1 \\
Pd    &  0.35  &  1.63  &  0.93  &  1.1 \\
\hline

\end{tabular}

Now we can see, that for all metals under consideration here, with the 
exception of Cu, experimental data on coefficient $\alpha$ are described rather
satisfactory and full agreement can be achieved by minor variations of 
$\lambda_{ee}$ around the value of $\lambda_{ee}=1$. This is not so only for Cu, 
where the agreement can be reached by introduction of $\lambda_{ee}<0$. 
The negative values of $\lambda_{ee}$ are possible, taking into account that 
the general limitation here \cite{PinNoz} is $F^s_1>-3$, i.e. $\lambda_{ee}>-1$.

Thus we came to the main conclusion -- the ``experimentally'' observed 
universal Planckian relaxation in metals, independent of the value of the
coupling constant, is nothing more than {\em delusion}, related to the
procedure of representation of experimental data in Refs. \cite{Bruin,Legros}
(determination of the effective mass in the expression for resistivity from
low -- temperature measurements). Similarly, the same applies to the results
of a recent paper  \cite{Cao}, where this procedure was used to represent
experimental data on bilayer graphene near the ``magic'' angle of (mis)orientation
of the layers.

\subsubsection{Once again on uncertainty principle}

Where is the problem with the estimates of relaxation time based on energy --
time uncertainty relation? This is rather simple -- in Eq.  (\ref{Heis}) 
$\Delta E$ should be taken not as $k_BT$, but as a real ``level width''
(spectrum damping) in the system of many interacting (!) particles:
\begin{equation}
\Gamma(T)=2\pi\lambda T
\label{Plkr}
\end{equation}
Then we immediately have:
\begin{equation}
\frac{\hbar}{\tau}\leq 2\pi\lambda T
\label{relax}
\end{equation}
so that the use of uncertainty relation becomes a kind of tautology.
Then the ``upper limit'' of high -- temperature relaxation rate is not
universal and naturally proportional to the coupling constant.

At the same time, the result of Eq. (\ref{PlanckGamma}) obtained above, 
formally defines some universal upper limit for relaxation rate of
{\em quasiparticles}, so that the estimates based on uncertainty relation
probably can be applied here.

\section{Conclusions}

As we stressed above, our aims did not included the explanation of $T$ --
linear growth of resistivity, starting from the lowest temperatures, which
is observed in copper oxides high -- temperature superconductors and some
similar systems. In the models of scattering by quantum fluctuations of an
arbitrary nature (e.g. phonons), such behavior appears at temperatures of the
order or higher than the characteristic frequency of these fluctuations, i.e.
in the classical limit. At present it is unclear whether such low -- frequency
fluctuations exist in the vicinity of the quantum critical point on the phase
diagram of cuprates, which could have explained their anomalous behavior.
But this a simplest calculable model which clearly shows, that the universal
Planckian behavior of relaxation rate is just absent, as well is just absent is
the universal upper limit for this relaxation rate. Dependence on the
interaction parameters in relaxation rate does not disappear and it can, in
principle, easily overcome the limit of  $\alpha\sim 1$.

In the literature, quite a number of microscopic models were proposed,
explaining the $T$ -- linear growth of resistivity (inelastic scattering rate)
in metals. We have already mentioned Refs. \cite{Khod,Khodel,Vol}, where this
behavior appeared in the model of ``fermion'' condensation and was related,
in particular, to formation of ``flat'' bands near the Fermi level and 
appearance of low -- frequency zero -- sound excitations, so that scattering by 
these excitations leads to $T$ -- linear growth of scattering rate.
In Ref. \cite{Sachdev} an interesting model was proposed with random
interactions of electrons, where the ``flat'' band also forms and for the wide
interval of parameters the $T$ -- linear growth of resistivity is realized.
It is unclear however, which relation if any at all the model of interactions
assumed in Ref.  \cite{Sachdev} has to real metals. Note also, that in all
these approaches the universal ``Planckian'' dependence on temperature does not
appear, and dependence on the coupling strength (though probably a weak one) 
is conserved. Similar situation is realized in a popular phenomenological
model of marginal Fermi -- liquid \cite{Varma}, which successfully explains
many properties of cuprates.

An interesting model was proposed recently in Ref. \cite{Mous} to explain the
anomalous temperature dependence of resistivity in Sr$_3$Ru$_2$O$_3$ by
electron -- electron scattering processes, taking into account rather complicated
real electronic spectrum of this compound, leading to many -- sheets of the
Fermi surface, with some special ``hot'' pockets, where electron are non --
degenerate (classical limit). It is the scattering of electrons from
``cold'' (degenerate) parts of the Fermi surface, with transition of one of
these electron to the ``hot'' pocket, which leads to the linear growth of resistivity
with temperature. The analogy with the model with scattering by non -- degenerate
fluctuations (classical limit) considered above is pretty obvious.
Of course, the dependence on the value of the appropriate coupling constant 
does not disappear, and the ``universal'' Palnckian behavior appears only for
the renormalized (in a sense discussed above) scattering rate of the 
quasiparticles.

The special place is occupied by Refs. \cite{Zan,Hart,Herz,Hartn}, based on the
analogies taken from black hole physics, cosmology and superstring theory, which
pretend to explain the universal ``Planckian'' behavior of the relaxation rate.
Being very interesting from theoretical point of view these papers, in our
opinion, have a weak relation to solid state physics.

\begin{figure}
\includegraphics[clip=true,width=0.45\textwidth]{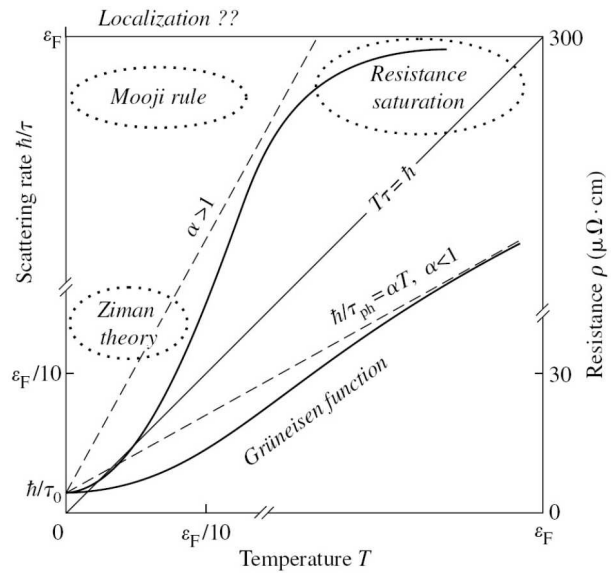}
\caption{Temperature -- disorder plane ($T$,$\hbar/\tau$), where regions of
applicability of different models of electrical resistivity are shown.
Also shown are temperature dependencies of resistivity. ``Planckian''
relaxation rate determines the diagonal on this plane.\cite{Gant}}
\label{Gantmaher}
\end{figure}

In this paper we have limited ourselves to elementary analysis based on the 
standard approaches of the solid state theory and shown, that the observed
universal ``Planckian'' behavior of electron relaxation rate in many metals
\cite{Bruin,Legros} has rather simple explanation, related not to some
special deep physics, but simply to the method of representation of
experimental data used in these papers. The use in Drude expression for 
conductivity (resistivity) of the effective mass, determined from low --
temperature measurements, which includes the renormalization due to many --
particle effects (interactions), inevitably leads to appropriate renormalization
of relaxation rate (time), which is replaced by an effective rate 
$\tilde\Gamma$ (\ref{tildaGamma}) of relaxation for quasiparticles, which is
relatively weakly dependent on the coupling strength and, in principle, 
produces the universal ``Planckian'' behavior, as an upper limit for this 
relaxation rate. However, this behavior is a pure {\em delusion} and does not
reflect any kind of special physics. In microscopic theory it is the
unrenormalized (by interactions) band mass, which enters the Drude expression
for conductivity, and relaxation rate is naturally proportional to the
strength of interaction, as was well -- known since the early days of quantum
solid state theory.

The size of an effective relaxation rate $\hbar/\tau=k_BT$
(i.e. in fact the ``Planckian'' rate discussed above) in a standard theory,
as is also well -- known, defines some characteristic scale, which separates
regions of different behavior of resistivity of metals, as it is shown in
Fig. \ref{Gantmaher}, taken from the textbook \cite{Gant}. There we also show
characteristic temperature dependencies of resistivity of metals.
``Planckian'' scattering rate determines the diagonal on this figure, but does 
not define any new quantum limit . In principle, all this is known for a long 
time and does not require any ``exotic'' approaches for its explanation.

The author is grateful to E.Z. Kuchinskii and D.I. Khomskii for useful 
discussions. This work was partially supported by RFBR grant 
No. 20-02-00011.


\appendix

\section{Momentum distribution in case of Planckian relaxation}

\footnotesize

Consider the momentum distribution corresponding to Green's function 
(\ref{MatGrP}):
\begin{equation}
N({\bf p})=T\sum_{n}G(i\varepsilon_n,{\bf p})
\label{distribution}
\end{equation}
Performing the standard summation over fermion Matsubara frequencies 
\cite{Diagram}, we obtain:\footnote{Calculations presented below were done by
E.Z. Kuchinskii}
\begin{eqnarray}
N({\bf p})=-\frac{1}{2\pi i}\int_{c}d\varepsilon f(\varepsilon)
G(\varepsilon,{\bf p})=\nonumber\\
-\frac{1}{2\pi i}\int_{-\infty}^{\infty}d\varepsilon f(\varepsilon)
(G^R(\varepsilon,{\bf p})-G^A(\varepsilon,{\bf p}))=
\int_{-\infty}^{\infty}d\varepsilon f(\varepsilon)A(\varepsilon,{\bf p})
\label{dist_general}
\end{eqnarray}
where $f(\varepsilon)=\frac{1}{e^{\varepsilon /T}+1}$ is Fermi distribution, 
and spectral density 
$A(\varepsilon,{\bf p})=-\frac{1}{\pi}ImG^R(\varepsilon,{\bf p})$ for Green's
fuction (\ref{MatGrP}) is given by Lorentzian:
\begin{equation}
A(\varepsilon,{\bf p})=\frac{1}{\pi}\frac{\frac{\alpha}{2}T}
{(\varepsilon-\varepsilon_{\bf p})^2+(\frac{\alpha}{2}T)^2}
\label{sp_dens}
\end{equation}

In Fig. \ref{planck_dist} we show distribution functions $N({\bf p})$, obtained
numerically directly from (\ref{dist_general}) with spectral density
(\ref{sp_dens}) for different values of $\alpha$. However, for special
$\alpha=2\pi n$ $n=0,1,\cdots$, distribution functions can be obtained
analytically. For $\alpha=0$ we obviously get the usual Fermi distribution
$N({\bf p})=f(\varepsilon_{\bf p})$. For $\alpha=2\pi$ we obtain:
\begin{equation}
G(i\varepsilon_n,{\bf p})=\frac{1}{i\varepsilon_n-\varepsilon_{\bf p}+i\pi T
sign\varepsilon_n}=
\frac{1}{i2\pi Tm-\varepsilon_{\bf p}}
\label{G_alpha2pi}
\end{equation}
where $m$ is an integer, but $sign\varepsilon_n$ excludes the value of $m=0$.
Then the distribution function of particles is:
\begin{eqnarray}
N({\bf p})=T\sum_{n}G(i\varepsilon_n,{\bf p})=T\sum_{m\neq 0}\frac{1}{i2\pi Tm-
\varepsilon_{\bf p}}=\nonumber\\
=T\sum_{m}\frac{1}{i2\pi Tm-\varepsilon_{\bf p}}+\frac{T}{\varepsilon_{\bf p}}
\label{n_alpha2pi_0}
\end{eqnarray}
Note that the sign before summation over seemingly even Matsubara frequencies
remained the same (fermion -- like), which is related to initial summing over
odd frequencies. The second term in the last expression in (\ref{n_alpha2pi_0}) 
compensated the contribution of $m=0$, which appeared in sum over $m$ in the
first term. Making the standard summation over even frequencies
$\varepsilon_m=2\pi Tm$ we get:
\begin{eqnarray}
&& T\sum_{m}\frac{1}{i\varepsilon_m-\varepsilon_{\bf p}}=
\frac{1}{2\pi i}\int_{c}d\varepsilon\frac{1}{\varepsilon-\varepsilon_{\bf p}\pm
i\delta}n(\varepsilon)=
\nonumber\\
&& =\frac{1}{\pi}\int_{-\infty}^{\infty}d\varepsilon Im\frac{1}{\varepsilon-
\varepsilon_{\bf p}+i\delta}n(\varepsilon)=
-\int_{-\infty}^{\infty}d\varepsilon\delta(\varepsilon-\varepsilon_{\bf p})
n(\varepsilon)=\nonumber\\
&& =-n(\varepsilon_{\bf p})
\label{sum_m}
\end{eqnarray}
where $n(\varepsilon)=\frac{1}{e^{\varepsilon /T}-1}$ is Bose distribution.

Thus, for $\alpha=2\pi$:
\begin{equation}
N({\bf p})=\frac{T}{\varepsilon_{\bf p}}-n(\varepsilon_{\bf p})
\label{n_alpha2pi}
\end{equation}
Note once again, that the minus sign before Bose distribution in 
(\ref{n_alpha2pi}) is related to initial summation over fermion frequencies.

\begin{figure}
\includegraphics[clip=true,width=0.45\textwidth]{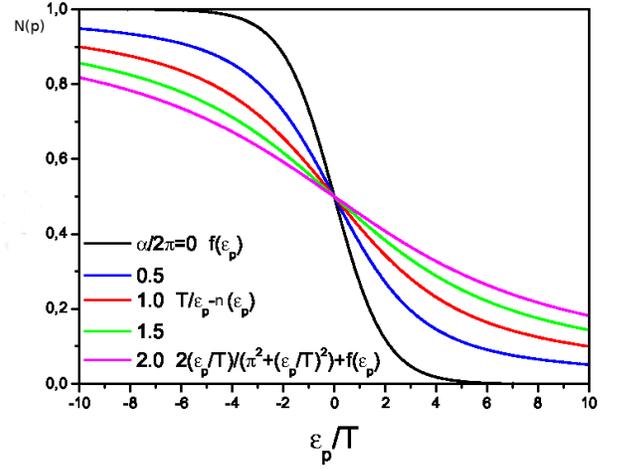}
\caption{Distribution function for Planckian relaxation with different $\alpha$.}
\label{planck_dist}
\end{figure}

For $\alpha=4\pi$ again we have summation over odd frequencies, but with two
closest to zero fermion Matsubara frequencies excluded:
\begin{eqnarray}
&& N({\bf p})=T\sum_{n}\frac{1}{i\varepsilon_n-\varepsilon_{\bf p}+i2\pi Tsign\varepsilon_n}=
\nonumber\\
&& =T\sum_{n\neq 0,-1}\frac{1}{i2\pi T(n+\frac{1}{2})-\varepsilon_{\bf p}}=
\nonumber\\
&& =T\sum_{n}\frac{1}{i\varepsilon_n-\varepsilon_{\bf p}}-
T\left(\frac{1}{i\pi T-\varepsilon_{\bf p}}+\frac{1}{-i\pi T-\varepsilon_{\bf p}}\right)=
\nonumber\\
&& =f(\varepsilon_{\bf p})+\frac{2T\varepsilon_{\bf p}}{\varepsilon_{\bf p}^2+(\pi T)^2}
\label{n_alpha4pi}
\end{eqnarray}
In Fig. \ref{planck_dist} for $\alpha=2\pi$ and $\alpha=4\pi$ we show both
distributions obtained numerically from (\ref{dist_general}) with spectral
density (\ref{sp_dens}) and obtained directly from (\ref{n_alpha2pi}) and 
(\ref{n_alpha4pi}) (obviously these are identical).

In general, it can be seen that distribution fictions in our model are rather
different from the usual step -- function of Fermi -- liquid theory.

{\bf Note added in proofs:} After this work was accepted for publication we have
learned on the paper by E.H. Hwang, S. Das Sarma. Phys. Rev. B{\bf 99}, 085105
(2019), where the authors came to the same conclusions in the context of
electron -- phonon interaction.

\end{document}